\documentclass[fleqn,usenatbib,letters]{mnras}

\usepackage{mathptmx}

\usepackage[T1]{fontenc}
\usepackage{ae,aecompl}

\usepackage{graphicx}	
\usepackage{amsmath}	
\usepackage{amssymb}	

\usepackage{threeparttable}
\usepackage{caption}
\usepackage{multirow}
\usepackage{arydshln}

\usepackage{textcomp}

\title[Lightning climatology of exoplanets]{Lightning climatology of exoplanets and brown dwarfs guided by Solar System data}

\author[G. Hodos\'an et al.]{G. Hodos\'an$^{1}$\thanks{E-mail:
gh53@st-andrews.ac.uk}, Ch. Helling$^{1}$, R. Asensio-Torres$^{2,3}$, I. Vorgul$^{1}$ and P. B. Rimmer$^{1}$\\
$^{1}$SUPA, School of Physics and Astronomy, University of St Andrews, St Andrews KY16 9SS, UK \\
$^{2}$Department of Astronomy, Stockholm University, Alba Nova University Center, SE-106 91 Stockholm, Sweden \\
$^{3}$Universidad de La Laguna, Departamento de Astrof\'isica, E-38206 La Laguna, Tenerife, Spain
}

\date{Accepted 2016 June 28. Received 2016 June 28; in original form 2015 October 19}

\pubyear{2016}

\begin{document}
\label{firstpage}
\pagerange{\pageref{firstpage}--\pageref{lastpage}}
\maketitle

\begin{abstract}
Clouds form on extrasolar planets and brown dwarfs where lightning could occur. Lightning is a tracer of atmospheric convection, cloud formation and ionization processes as known from the Solar System, and may be significant for the formation of prebiotic molecules. We study lightning climatology for the different atmospheric environments of Earth, Venus, Jupiter and Saturn. We present lightning distribution maps for Earth, Jupiter and Saturn, and flash densities for these planets and Venus, based on optical and/or radio measurements from the WWLLN and STARNET radio networks, the LIS/OTD satellite instruments, the \textit{Galileo}, \textit{Cassini}, \textit{New Horizons} and \textit{Venus Express} spacecraft. We also present flash densities calculated for several phases of two volcano eruptions, Eyjafjallaj\"okull's (2010) and Mt Redoubt's (2009). We estimate lightning rates for sample, transiting and directly imaged extrasolar planets and brown dwarfs. Based on the large variety of exoplanets, six categories are suggested for which we use the lightning occurrence information from the Solar System. We examine lightning energy distributions for Earth, Jupiter and Saturn. We discuss how strong stellar activity may support lightning activity. We provide a lower limit of the total number of flashes that might occur on transiting planets during their full transit as input for future studies. We find that volcanically very active planets might show the largest lightning flash densities. When applying flash densities of the large Saturnian storm from 2010/11, we find that the exoplanet HD 189733b would produce high lightning occurrence even during its short transit.
\end{abstract}





\section{Introduction} \label{sec:int}

Since 1995, the discovery of the first exoplanet around a Sun-like star \citep{mayor1995}, a wide variety of exoplanets have been observed, including hot Jupiters (e.g. HD 189733b), mini-Neptunes (e.g. Kepler-11c), super-Earths (e.g. 55 Cnc e) and even planets smaller than Earth (e.g. Kepler-70c). The different techniques used for detection allow exploring these extrasolar objects from different points of view. Radial-velocity measurements and transit observations together give a constraint on the radius and mass of the planet. Transmission spectroscopy reveals information regarding the planetary atmosphere. The orbit of the planet can be mapped with direct imaging, and microlensing could map the frequency of different sized planets around different stars in the Galaxy disc, since it is not biased towards certain stellar or planetary types. In this paper we will focus on brown dwarfs, and planets discovered by either the transit method or direct imaging. 

Planets analysed through transit spectroscopy are observed to have clouds in their atmospheres, most likely made of silicate particles \citep[e.g.][]{kreidberg2014, sing2009, sing2013, sing2015}. These findings are supported by kinetic cloud models as in \citet{helling2008b, helling2011, helling2011b}. Various authors demonstrated that atmospheric circulation leads to the formation of zonal jets and local vortices as known from Jupiter and Saturn \citep[e.g.][]{dobbsdixon2012, dobbs-dixon2013, mayne2014}. E.g. \citet{zhang2014} found that strong internal heating and weak radiative dissipation results in the formation of large-scale jets. \citet{lee2015} modelled local and global cloud patterns on the planet HD 189733b, a tidally locked hot Jupiter orbiting a K star. Their dust opacity, grain size distribution and albedo maps indicate that cloud properties change significantly from dayside to night side forming a spot-like cloud pattern driven by a latitudinal wind jet around the equator.

Clouds in dynamic atmospheres are commonly associated with lightning. On Jupiter and Saturn, lightning is produced in dense, vertically extended, convective clouds \citep{dyudina2001, dyudina2004, dyudina2013, read2011}. Lightning on Venus may appear as intracloud (inside a cloud deck, IC) or intercloud (between clouds) discharge due to the high atmospheric pressure, which would not allow cloud-to-ground (CG) discharges to occur unless the electric field becomes extremely high \citep{yair2008}. Lightning observations of the Solar System planets apply methods of combined detection of optical and radio signals, which are well tested for lightning detection on Earth. \citet{helling2013} showed that lightning can be expected in extrasolar planetary atmospheres. \citet{vorgul2016} suggested that present day radio observations of brown dwarfs may contain hints to the presence of lightning in these atmospheres.

Dedicated observational campaigns have revealed that lightning occurs in very diverse environments on Earth. Lightning is frequently produced in thunderclouds that are made of water and ice particles. Thunderstorms also occur in clouds of ice and snow particles, producing "winter lightning" \citep{brook1982, wu2013}. During explosive volcanic eruptions intense lightning activity is observed in volcano plumes, which are primarily composed of mineral dust particles \citep{james2008}. 
Lightning has been suggested as a tool to study, for example, earthquake occurrence, and a relation with global warming was indicated. \citet{mullayarov2007} investigated the relation between lightning radio signatures originating from thunderstorms passing over earthquake regions and earthquake activity. \citet{romps2014} suggested a link between global warming over the United States and flash rate variability. Their results suggest an increase of flash numbers due to an increase of global precipitation rate and of the convective available potential energy (CAPE), a proxy of lightning activity. This suggests that lightning in the astrophysical context  will depend on internal heating and stellar irradiation that will affect the local atmospheric temperature, which determines where clouds form. Consequently, lightning activity on a planet will be affected by the age, and hence the magnetic activity of the host star and by the planet's distance from the star. In the case of brown dwarfs, it is the age of the object that counts most as this determines its total energy household including magnetic activity driven by rotation. If the brown dwarf resides in a binary system \citep[e.g.][]{casewell2012, casewell2013, casewell2015}, the characteristics of the companion may also play a role in the production of lightning discharges. \citet{desch2000} proposed that lightning in the solar nebula was the main cause of the presence of chondrules, millimetre-sized glassy beads, within meteorites. \citet{nuth2012} suggested that lightning may play an important role in the evolution of oxygen isotopes in planetary discs. They suggested that lightning activity in stellar nebulae may affect the $^{16}$O and $\delta^{17}$O production, which could lead to the observed, but not yet fully explained, non-equilibrium appearance of $\delta^{17}$O and $\delta^{18}$O isotopes in primitive meteorites. 

This paper presents an analysis of lightning surveys on Earth, Venus, Jupiter and Saturn, as lightning detection efforts were focused on these planets. Our planetary system provides opportunities to compare different environments where lightning occurs, and therefore, provides guidance for the large diversity of exoplanets and their atmospheres. We compare lightning climatology from these Solar System planets and use these statistics as a guide for a first consideration of lightning activity on extrasolar objects. We use lightning climatology maps to find patterns in the spatial distribution of lightning strikes, such as increased lightning activity over continents than over oceans, and calculate flash densities (flashes km$^{-2}$ year$^{-1}$ and flashes km$^{-2}$ hour$^{-1}$) and flash rates (flashes unit-time$^{-1}$) in order to estimate the total number of events at a certain time over a certain surface area. Estimating the number of lightning flashes and their potential energy distribution is essential for follow-up studies such as lightning chemistry \citep[e.g.][]{rimmer2016} in combination with 3D radiative hydrodynamic models \citep{lee2016}. 

The paper is organized in three main parts. Section \ref{sec:earth} summarizes  Earth lightning observations, in the optical (direct lightning detection) and radio (low frequency (LF) emission) data taken by several Earth-based stations (STARNET, WWLLN) and Earth-orbiting satellites (OTD/LIS), and compares the data by exploring the detection limits, general trends and differences between the data sets. In Sect. \ref{sec:solsys}, we explore lightning observations on Venus, Jupiter and Saturn by summarizing and analysing data from various spacecraft and creating lightning maps. In Section \ref{sec:exopl}, we use the lightning climatology data as guide for potential lightning occurrence on the diverse population of exoplanets. Specific exoplanets are discussed and brown dwarfs are also included in this section.
Section \ref{sec:con} summarizes this paper.


\section{Lightning data from Earth} \label{sec:earth}

\begin{table*} 
\begin{threeparttable}
 \small 
 \centering
 \begin{minipage}{110mm}
 \caption{Properties of instruments used for lightning detection on Earth. (FoV = Field of View.) OTD: \citet{boccippio2000, boccippio2002, beirle2014}. LIS: \citet{christian2003, cecil2014, beirle2014, christian2000}. STARNET: \citet{morales2014}. WWLLN: \citet{abarca2010, hutchins2012, hutchins2013}.}
  \begin{tabular}{@{}lccllc@{}}	
	\hline
	Instrument/Network & Spatial resolution & Temporal resolution & Detection threshold & FoV/Coverage & Detection Efficiency \\
	\hline
	OTD & $10 - 11$ km & $\sim 2$ ms & 9-21 $\mu$J m$^{-2}$ sr$^{-1}$ & 1300 km $\times$ 1300 km\tnote{(1)} & \vtop{\hbox{\strut day: 40\%}\hbox{\strut night: 60\%}} \\
	LIS & $4 - 6$ km & 2 ms & 4-11 $\mu$J m$^{-2}$ sr$^{-1}$ & 600 km $\times$ 600 km & \vtop{\hbox{\strut day: 70\%}\hbox{\strut night: 90\%}} \\
	STARNET & $5-20 $ km & 1 ms\tnote{(2)} & - (no information) & \vtop{\hbox{\strut South America}\hbox{\strut Caribbean}\hbox{\strut SW-Africa}} & \vtop{\hbox{\strut day: 45\%}\hbox{\strut night: 85\%}}\\
	WWLLN & $\sim 5$ km & $\sim 15 \mu$s & \vtop{\hbox{\strut Space, time and}\hbox{\strut station dependent}} & Full Earth & $\sim 2-13$\% \\
	\hline
  \label{table:instr}
  \end{tabular}
  \begin{tablenotes}
	\item[1] http://thunder.msfc.nasa.gov/otd/  
	\item[2] http://www.starnet.iag.usp.br/index.php?lan=en     
  \end{tablenotes}
 \end{minipage}
\end{threeparttable}
\end{table*}

Earth is the most well-known planet we can learn from and apply as an analogue for exoplanetary sciences. Both observational and theoretical works that used Earth as a guide have been conducted to analyse different features of exoplanets. \citet{palle2009}, for example, compared the transmission spectrum of Earth taken during a lunar eclipse and the spectrum of the Earthshine, which is the reflection spectrum of Earth. They used the transmission spectrum as an analogue for a primary transit of Earth as seen from outside the Solar System, while the reflection spectrum is an indicator of a directly imaged exo-Earth after removal of the Sun's features. Similar studies of Earth as an exoplanet, such as looking for vegetation or other signatures caused by biological activity, were conducted by e.g. \citet{montanes2010, arnold2002, sterzik2009, kaltenegger2007}.

Lightning detection and statistics on Earth are very important because of the hazards (e.g. forest fires, large scale power outage, fatalities) it causes. Lightning detecting networks are set up on the surface of the planet while satellites monitor the atmosphere for lightning events. Earth measurements provide the largest data set due to the continuous observations and the high spatial coverage of the instruments. Data used in this paper were provided by the \textit{Lightning Imaging Sensor} (LIS)/\textit{Optical Transient Detector} (OTD) instruments on board of satellites in the optical, and two ground based radio networks, the \textit{Sferics Timing and Ranging Network} (STARNET) and \textit{World Wide Lightning Location Network} (WWLLN). WWLLN and STARNET detect strokes\footnote{events with discrete time and space} while LIS/OTD observe flashes\footnote{events with duration and spatial extent; one flash contains multiple strokes} \citep{rudlosky2013}. A more detailed description of the instruments and the obtained data can be found in Appendix A1 and A2. Table \ref{table:instr} lists relevant properties of the lightning detecting instruments and networks.


\subsection{Detection efficiency} \label{subs:de}

The detection efficiency (DE) is the detected percentage of the true number of flashes \citep{chen2013}. It depends on the sensitivity threshold of the instrument, geographic location, and time of the observation \citep{cecil2014}. Seen from an astronomical perspective, the DE is extremely well determined for Earth, however, less so for the Solar System planets. Therefore, we use the knowledge from Earth to discuss the impact of the DE on the lightning data, in order to understand the limits, but also the potentials of the available data for exoplanetary research.

For LIS/OTD the DE is determined by two different approaches. \citet{boccippio2000} cross-referenced individual flash detections with the U.S. National Lightning Detection Network data, which provides an empirical estimate on the DE. \citet{boccippio2002} used independent measurements of pulse radiance distributions to model the DE. The estimated DEs for OTD and LIS are listed in Table \ref{table:instr}.

The DE for STARNET is determined by comparing detections with other networks (e.g. with World Wide Lightning Location Network (WWLLN)). According to the comparison studies conducted by \citet{morales2014} STARNET detects $\sim 70\%$ of lightning strokes, however this value depends on the antennas in use and it has a diurnal pattern (85\% day, 45\% night DE). Two different WWLLN DEs are quoted in the literature: relative DE (RDE) and absolute DE (ADE). The RDE is determined by the model given in \citet{hutchins2012} that is based on the detected energy per stroke: once the energy distribution of observed samples is known, the missing energies (and amount of lightning) can be estimated. The RDE compensates for the uneven distribution of sensors on Earth and variations in very low frequency (VLF) radio propagation and allows representing the global distribution of strokes as if it was observed by a globally uniform network \citep{hutchins2012}. The ADE was determined by comparing WWLLN data with other networks. \citet{abarca2010} cross-correlated stroke locations with detections of the National Lightning Detection Network (NLDN) data and found that WWLLN DE is highly dependent on the current peak and polarity of the lightning discharge and varies between $\sim 2 - 11 \%$. \citet{rudlosky2013} showed the improvement of WWLLN DE between 2009 and 2013 compared to LIS observations (up to $\sim 10 \%$), while \citet{hutchins2012} found the ADE to be $\sim 13$\%. In our calculations, following \citet{rudlosky2013}, we took the WWLLN's DE to be 9.2\% for 2012 under the assumption that LIS was 100\% efficient. 

The DE is an important parameter of the lightning detecting instruments, however, it cannot be determined perfectly and unambiguously. It introduces an uncertainty in the measurements, it is estimated based on models and/or comparison studies. Models include estimates (e.g. see the models of \citet{boccippio2002}), and comparison studies assume a lightning detecting network/satellite to be, ideally, 100\% efficient. Since the true value of the DE of an instrument or network is unknown, the obtained flash densities are only a lower limit of the total number of flashes occurring on Earth at a certain time. No DEs are yet available for the lightning observations on Venus, Jupiter and Saturn. Therefore, it seems justified to conclude that the Solar System data, including Earth, are a lower limit for lightning occurrence on these planets.


\subsection{Lightning climatology on Earth} \label{sec:e_data}

In this section, we derive and compare flash densities for the different networks and satellites based on already published, extensive data from Earth.

\begin{figure*}
  \includegraphics[width=\columnwidth, trim=0cm 0cm 0cm 0cm]{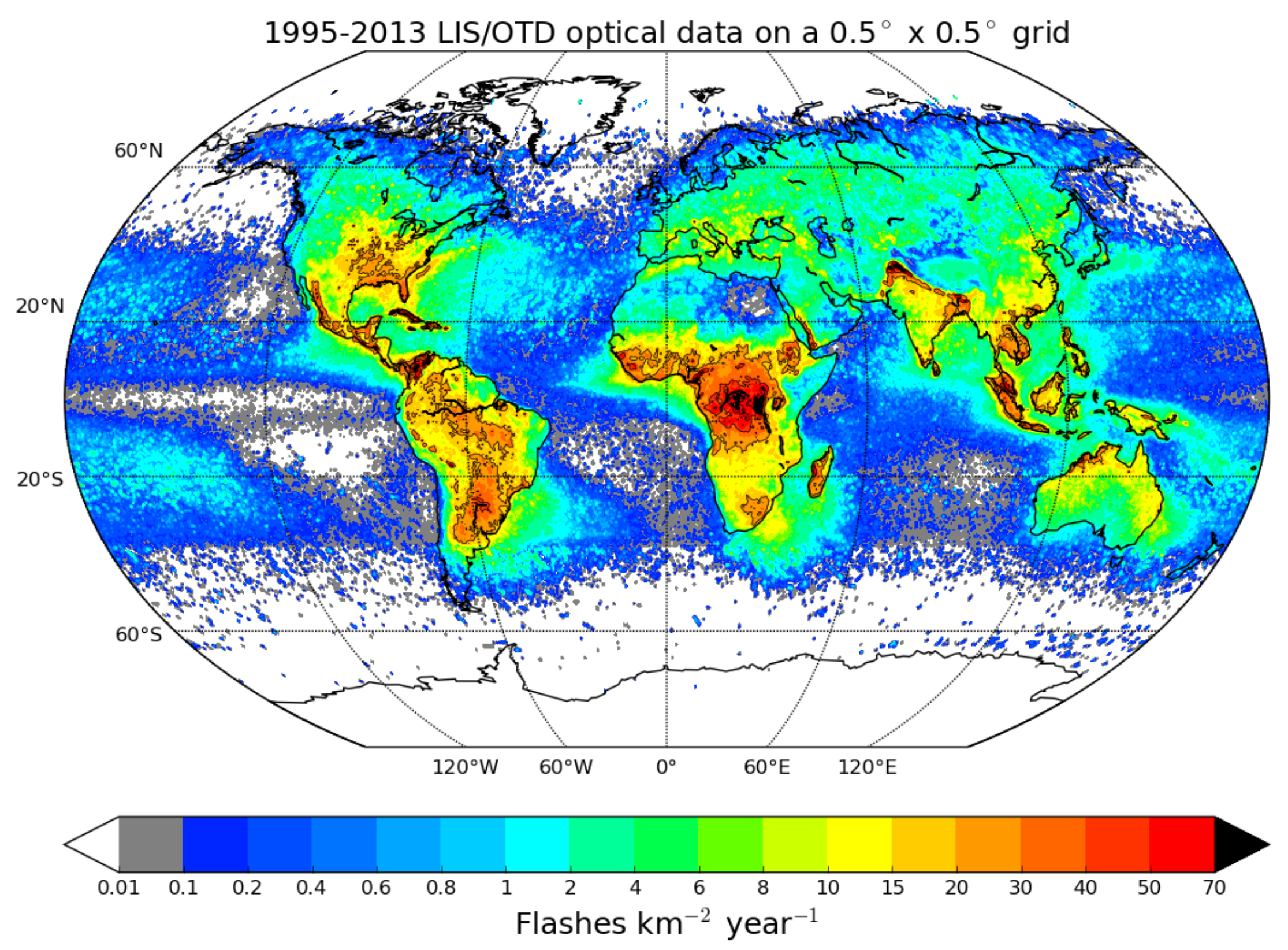}
  \includegraphics[width=\columnwidth, trim=0cm 0cm 0cm 0cm]{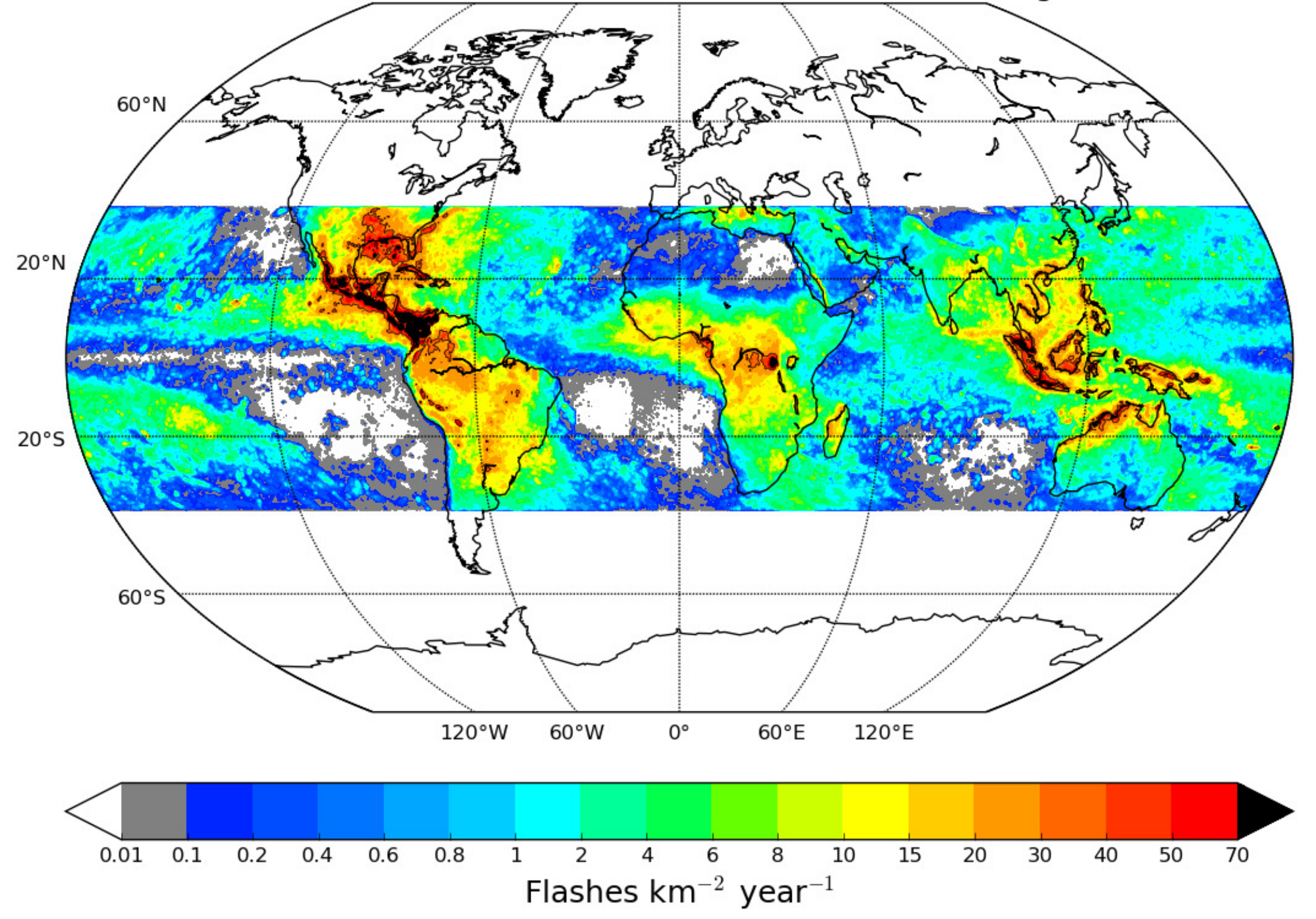}
  \caption{\textbf{Left:} Mean annual flash density from optical LIS/OTD data averaged on a $0.5$\textdegree$ \times 0.5$\textdegree geographical grid across Earth's surface \citep[for description of the data see section on "High resolution flash climatology (HRFC)" in][]{cecil2014}. LIS covers the area between $\pm 38$\textdegree in latitude and the years 1998-2013, while OTD monitored the whole globe (excluding polar regions) in the period of 1995-2000 \citep{cecil2014}. The map shows the differences between continents and oceans. Most of the lightning activity was recorded over continents, especially on low-latitudes.
\textbf{Right:} WWLLN mean annual flash density on a $0.5$\textdegree $\times 0.5$\textdegree grid across the LIS field of view. WWLLN data were scaled by DE and strokes were converted into flashes to match the LIS observations. Comparing it to the figure on the left, we find that WWLLN detects less flashes than LIS/OTD.}
  \label{fig:1}
\end{figure*}

\begin{figure*}
  \centering
  \includegraphics[width=\columnwidth, trim=0cm 0cm 0cm 0cm]{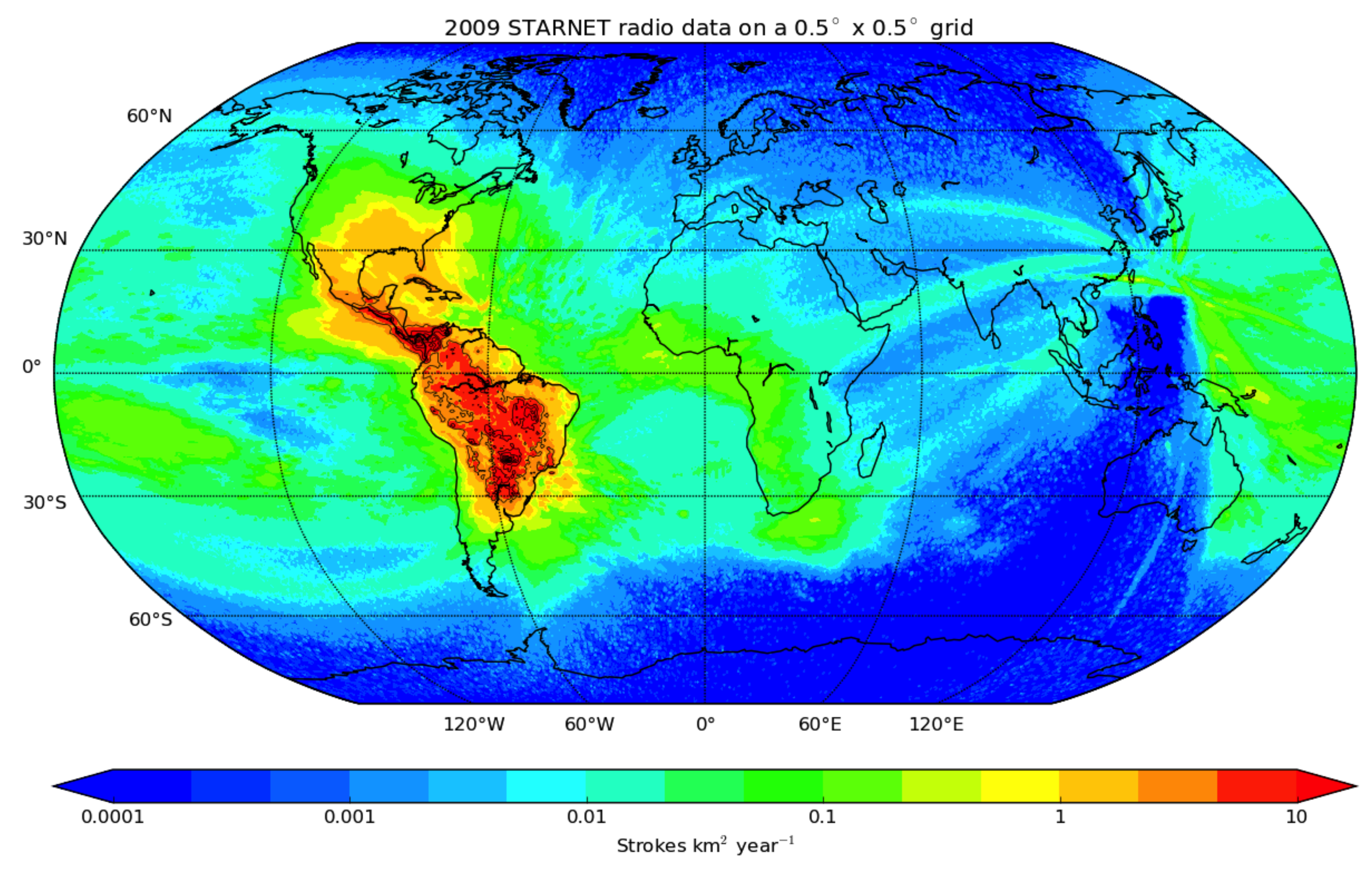}
  \includegraphics[width=\columnwidth, trim=0cm 0cm 0cm 0cm]{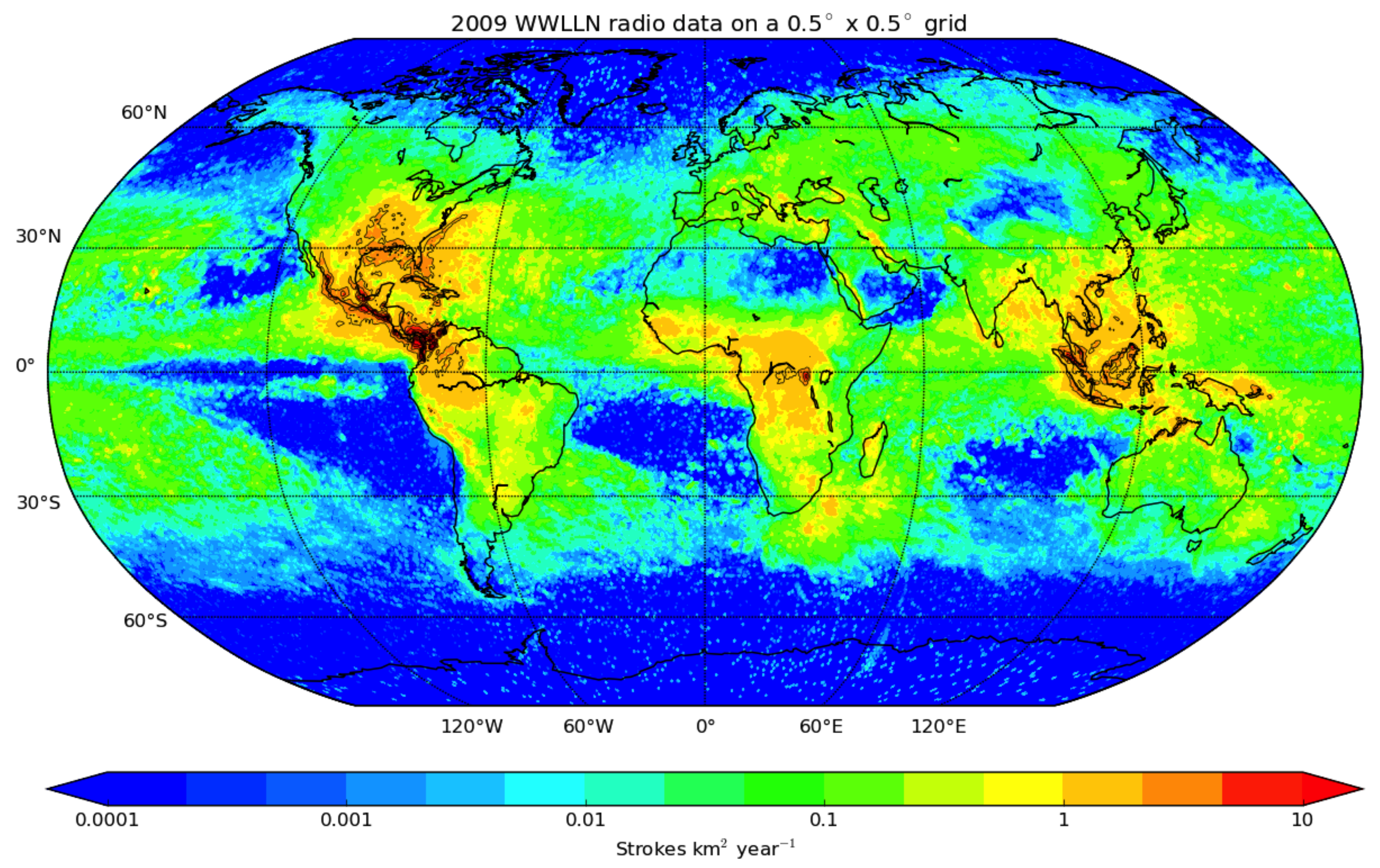} \\
  \includegraphics[width=\columnwidth, trim=0cm 0cm 0cm 0cm]{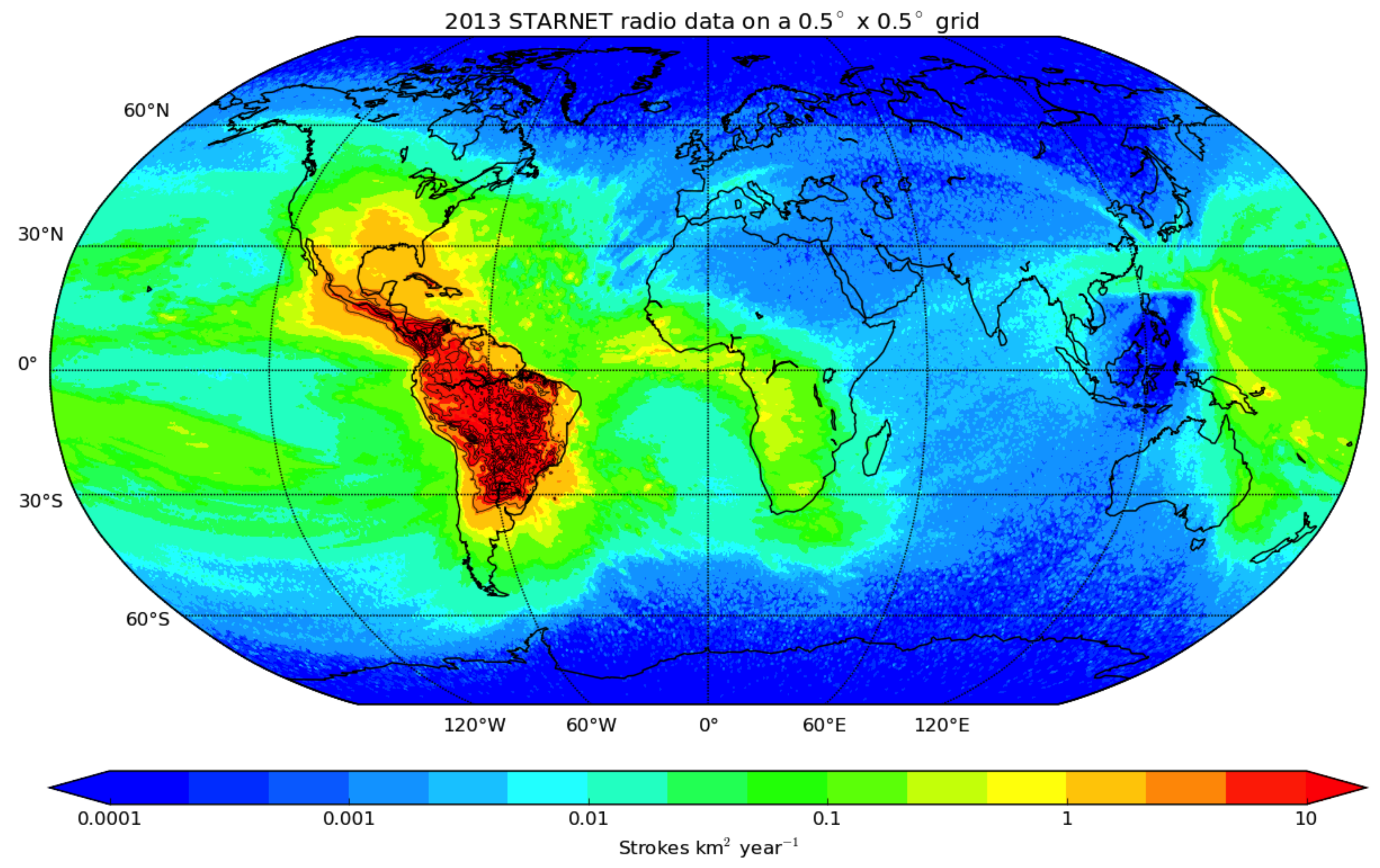} 
  \includegraphics[width=\columnwidth, trim=0cm 0cm 0cm 0cm]{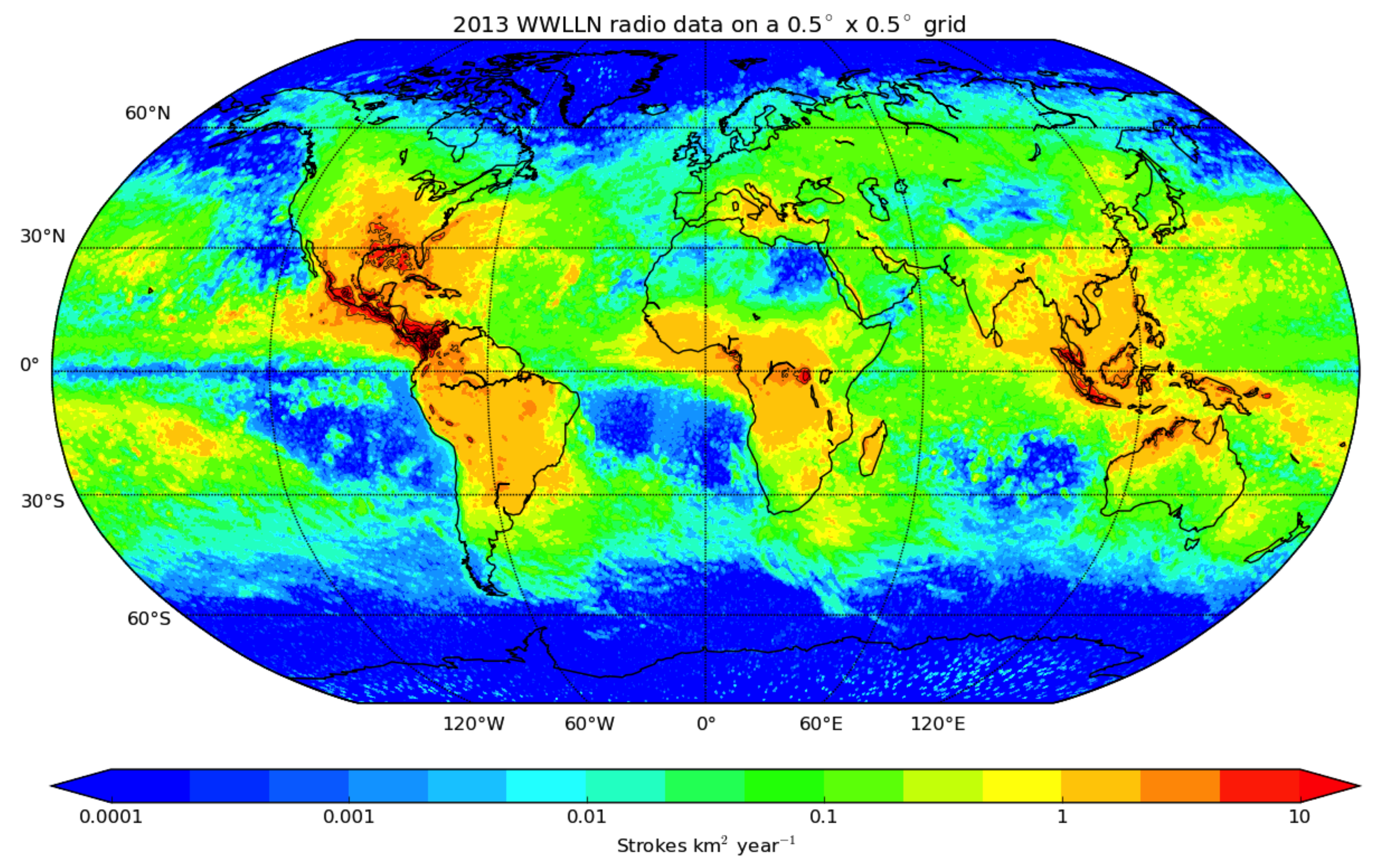}
  \caption{Mean annual stroke density on a $0.5$\textdegree $\times 0.5$\textdegree grid, created from very-low frequency (VLF) STARNET ($7 - 15$ kHz, left) and WWLLN ($3 - 30$ kHz, right) radio data. \textbf{Top:} 2009, \textbf{Bottom:} 2013. In 2009, the Sun was close to the minimum of its 11-year cycle, while 2013 was close to solar maximum. Comparing the two maps of the same instruments, they show more lightning in 2013 than in 2009. The STARNET data (left) show an arch-like trend above the Indian Ocean and Asia, which is, most probably, a numerical or observational artefact.
}
  \label{fig:4}
\end{figure*}

Figures \ref{fig:1}-\ref{fig:4} show flash densities averaged and plotted on a $0.5$\textdegree $\times 0.5$\textdegree geographical grid.
The left panel of Fig. \ref{fig:1} shows the mean annual flash densities \citep[flashes km$^{-2}$ year$^{-1}$,][]{cecil2014} based on LIS/OTD data in the period of 1995-2013. The LIS/OTD data show lower flash densities over oceans and dry regions than continents. Fewer flashes are detected at high latitudes (e.g. Canada, Siberia, etc.), than  at lower latitudes. \citet{cecil2014} derived the global average flash density from the $0.5$\textdegree $\times 0.5$\textdegree high resolution data set to be 2.9 flashes km$^{-2}$ year$^{-1}$ and the peak value to be 160 flashes km$^{-2}$ year$^{-1}$. We reproduce their results from the original data to be $\sim 2$ flashes km$^{-2}$ year$^{-1}$ for the annual average and $\sim 163$ flashes km$^{-2}$ year$^{-1}$ for maximum values.

The left column of Fig. \ref{fig:4} shows maps with annual stroke densities (strokes km$^{-2}$ year$^{-1}$) from STARNET data for the years 2009 (top) and 2013 (bottom). For these years STARNET had a coverage over the Caribbean, South America and western Africa. The right column of Fig. \ref{fig:4} shows the mean annual stroke density maps for 2009 (top) and 2013 (bottom) created from WWLLN data (missing 15 days from Apr 2009). WWLLN shows similar stroke distribution pattern to LIS/OTD, more lightning over continents than oceans, although WWLLN finds the maximum of lightning strokes (km$^{-2}$ year$^{-1}$) over Central-America, while LIS/OTD shows the most lightning over Africa (Fig. \ref{fig:1}, left).  

The effects of different DEs are seen in Figure \ref{fig:4} for STARNET and WWLLN data. If we choose one of the years, e.g. 2009, and focus on the South-American region, it is clearly seen that STARNET detects more strokes than WWLLN. STARNET operates more radio antennas in this region, than WWLLN, which increases the DE of the network. Data from two years (2009, 2013) are plotted in Figure \ref{fig:4}. The two years were chosen in order to represent different phases of solar activity: there was a solar minimum in 2009, while in 2013 the Sun was very active\footnote{http://www.climate4you.com/Sun.htm - Climate4you developed by Ole Humlum}. Comparing the data for the two years in Fig. \ref{fig:4} leads to the conclusion that more lightning strokes were observed in 2013 ($\sim$ solar maximum) than in 2009 ($\sim$ solar minimum). However, in case of WWLLN the increase of detected lightning strokes may be the reason of increased DE between 2009 and 2013 \citep{rudlosky2013}, hence the correlation with solar activity remains uncertain. (A more detailed comparison between solar activity and lightning activity is discussed in Sect. \ref{sec:stelact}.) The maps from the two years can be correlated with El Ni\~no events. El Ni\~no was observed in 2009, however not in 2013\footnote{http://ggweather.com/enso/oni.htm - by Jan Null}. Interestingly, Fig. \ref{fig:4} shows more lightning activity in 2013, on the contrary to what is expected from previous studies showing slightly larger lightning activity during El Ni\~no periods over tropical and sub-tropical continental regions \citep[e.g.][]{satori2009, siingh2011}.

WWLLN strokes were scaled by the DE and converted into flashes to match the LIS data by assuming 1.5 strokes/flash \citep{rudlosky2013}. The right panel of Fig. \ref{fig:1} demonstrates that WWLLN detects fewer flashes in Africa than LIS (Fig. \ref{fig:1}, left). This suggests that the difference between the detections is caused by the lower WWLLN DE in Africa. Flashes may contain more than 1.5 strokes \citep{rakov2003}, in which case the WWLLN would detect even less flashes than the LIS satellite.

The flash densities are summarized in Table \ref{table:plan} and their potential application to exoplanets is discussed in Sect. \ref{sec:exopl}.

\subsection{Lightning in volcano plumes} \label{sec:volc}

\begin{table*}
\begin{threeparttable}
 \small 
 \caption{Volcano eruptions investigated in this study, their characteristics and calculated lightning flash densities. We determined the flash densities based on reported observations as described in Sect. \ref{sec:volc}. The values are used to estimate lightning occurrence on the exoplanets Kepler-10b and 55 Cnc e, and the brown dwarf Luhman-16 B (Sect. \ref{sec:casest}).}
  \begin{tabular}{@{}lllllc@{}}	
	\hline 
	N$^{\rm o}$ & Volcano & Eruption date & Information & Reference & \vtop{\hbox{\strut Average flash densities}\hbox{\strut [flashes km$^{-2}$ hour$^{-1}$]}} \\
	\hline \hline
	$[1]$ & \multirow{6}{*}{Eyjafjallaj\"okull} & 14-19 Apr 2010 & \vtop{\hbox{\strut Electrically active for $\sim90$ h}\hbox{\strut 171 strokes observed}\hbox{\strut Standard deviation of location: 4.8 km}} & \multirow{6}{*}{\citet{bennett2010}} & 0.1 \\
	 $[2]$ & & 11-20 May 2010 & \vtop{\hbox{\strut Electrically active for $\sim235$ h}\hbox{\strut 615 strokes observed}\hbox{\strut Standard deviation of location: 3.2 km}} & & 0.32 \\ \hline
	$[3]$ & \multirow{5}{*}{Mt Redoubt} & 23 Mar 2009\tnote{(1)} & \vtop{\hbox{\strut Electrically active for 20.6 min}\hbox{\strut 573 flashes observed}\hbox{\strut Farthest sources from the vent: 28 km}} & \multirow{5}{*}{\citet{behnke2013}} & 12.04 \\ 
	 \vtop{\hbox{\strut $[4]$}\hbox{\strut $[5]$}} & & 29 Mar 2009\tnote{(1)} & \vtop{\hbox{\strut Phase 1: 100 flashes min$^{-1}$ per 3 km$^{2}$}\hbox{\strut Phase 2: 20 flashes min$^{-1}$ per 11 km$^{2}$}} & & \vtop{\hbox{\strut 2000.0}\hbox{\strut 109.0}}\\ 
	\hline
  \label{table:vol}
  \end{tabular}
  \begin{tablenotes}
	\item[1] One of the twenty-three episodes occurring in March-April 2009 \citep{behnke2013}    
  \end{tablenotes}
\end{threeparttable}
\end{table*}

Electrical activity has long been associated with large-scale, explosive volcanic eruptions \citep{james2008, mather2006}. There are records on lightning events from 1650, occurring at a volcanic eruption near Santorini, Greece \citep{fouque1879}. Eye-witnesses reported electrical phenomena, which coincided with the eruption of the Krakatoa in Indonesia in 1883 \citep{symons1888}. The modern era has produced a high number of volcanic lightning observations, after volcanic eruptions like, e.g., Etna in 1979, 1980; Mt St Helens in 1980, 1983; Gr\'imsv\"otn in 1996, 1998, 2004; or Hekla in 2000; etc. \citep[for references and an extended list see][]{mather2006}. 

In this section we analyse statistics from two volcanic eruptions: the Icelandic Eyjafjallaj\"okull's eruption from 2010 and the Mt Redoubt eruption in Alaska, 2009 (Table \ref{table:vol}). We derive flash densities (Table \ref{table:vol}), which we use to estimate lightning activity in rocky exoplanet and brown dwarf atmospheres (Sect. \ref{sec:exopl}). The composition of volcanic plumes may reflect the composition of dust clouds on these extrasolar objects.

The Eyjafjallaj\"okull eruption had two main phases: 14-19 April 2010, with 171 strokes occurring in about 90 hours, and 11-20 May 2010, a more intensive one with 615 strokes in about 235 hours \citep{bennett2010}. The standard deviation of the location of the lightning events was 4.8 and 3.2 km, respectively \citep{bennett2010}. We used this information to estimate the influenced area, assuming that the area is a circle with the diameter of the standard deviation. We calculate the stroke density for the two phases to be 0.1 strokes km$^{-2}$ h$^{-1}$ and 0.32 strokes km$^{-2}$ h$^{-1}$, respectively. \citet{bennett2010} measured the multiplicity of the flashes, the number of strokes occurring in one flash, and found that only 14 flashes had 2 strokes, while all other flashes were composed of single strokes. Based on this information, we assume that the flash densities during the 2010 Eyjafjallaj\"okull eruption are equal to the calculated stroke densities (14-19 April 2010: 0.1 km$^{-2}$ h$^{-1}$; 11-20 May 2010: 0.32 km$^{-2}$ h$^{-1}$).

\citet{behnke2013} analysed various episodes of the 2009 Mt Redoubt eruption. We used the information on two episodes: the 23 March 2009 episode, which resulted in the occurrence of 573 lightning flashes in 20.6 minutes (0.34 hours); and the 29 March 2009 episode with two main phases, the first with 100 flashes min$^{-1}$ over a 3 km$^2$ area and the second with 20 min$^{-1}$ over 11 km$^2$. During 23 March 2009 the farthest sources were located 28 km from the vent \citep{behnke2013}, which suggest that vent dynamics may not be the primary driver for this lightning. Assuming the affected area can be approximated by a rectangle of sizes 28 km $\times$ 5 km \citep[][fig. 6]{behnke2013}, the total affected area would be 140 km$^2$. The obtained average flash density for the 23 March 2009 episode is 12.04 km$^{-2}$ h$^{-1}$. The episode 29 March 2009 show much larger flash densities, with 2000 km$^{-2}$ h$^{-1}$ for the intensive first phase and 109 km$^{-2}$ h$^{-1}$ for the longer second phase.

\citet[][table 3]{mather2006} list flash densities based on \citet{anderson1965} for volcano plumes to be between 0.3 and 2.2 km$^{-2}$ min$^{-1}$, which is 18 and 132 km$^{-2}$ h$^{-1}$, respectively. The large lightning storm on 29 March 2009 around Mt Redoubt shows comparable flash densities during its second phase. The obtained flash densities (Table \ref{table:vol}) are used to estimate lightning occurrence on rocky exoplanets without water surfaces, and on brown dwarfs, since clouds on these types of objects may resemble volcano plumes. Also, we note that lightning statistics are not well studied in case of volcano eruptions. The values listed in Table \ref{table:vol} (last column) are guides and may only be used under certain assumptions as we outlined in Sections \ref{sec:casest} and \ref{sec:flashdens}.

\section{Lightning on other Solar System planets} \label{sec:solsys}

\subsection{Lightning on Venus?} \label{sec:venus}

The presence of lightning on Venus has been suggested by multiple observations since the late 1970s. \citet{ksanfomaliti1980} reported lightning detection based on the data gathered by the \textit{Venera 11} and \textit{Venera 12} landers. \citet{scarf1980} presented whistler\footnote{electromagnetic waves emitted in the VLF range, propagating through plasma along magnetic field lines \citep{desch2002}} detections by the \textit{Pioneer Venus Orbiter} (PVO). However, these early observations were not widely accepted. \citet{taylor1987}, for example, interpreted the VLF radio signals as interplanetary magnetic field/solar wind related perturbations appearing around the PVO spacecraft. Since then several attempts have been made to detect lightning on Venus, and the controversy of the existence of lightning on the planet has not yet been resolved. The \textit{Cassini} spacecraft made two close fly-bys of Venus in 1998-99, but did not detect lightning induced radio emission in the low frequency range \citep{gurnett2001}. \citet{gurnett2001} calculated a lower limit of flash rate from the non-detection to be 70 s$^{-1}$, slightly smaller than the average on Earth (100 s$^{-1}$). \citet{krasnopolsky2006} detected NO in the infrared spectra of Venus, which they related to lightning activity in the lower atmosphere of the planet and inferred a flash rate of 90 s$^{-1}$. Other attempts of optical observations were conducted by \citet{garcia2011} to observe the 777 nm O emission line (prominent signature of lightning in the Earth atmosphere), however no detection was reported, which suggest a rare Venusian lightning occurrence or at least that it is less energetic than Earth lightning. (For a summary of lightning observations on Venus see \citet{yair2008, yair2012}.)

In 2006, when the \textit{Venus Express} reached Venus, a new gate to lightning explorations opened \citep[e.g.][]{russell2008, russell2011, daniels2012, hart2014b, hart2014}. \citet{russell2008} reported whistler detections by \textit{Venus Express} near the Venus polar vortex from 2006 and 2007, which they associated with lightning activity and inferred a stroke rate of 18 s$^{-1}$. The MAG (Magnetometer) on board of \textit{Venus Express} detected lightning induced whistlers in 2012 and 2013 too. The data were analysed by \citet{hart2015}, who confirmed the whistler events with dynamic spectra. However, since the magnetic field around Venus is not yet fully understood, the field lines cannot be traced back to their origin, therefore, the coordinates of the source of the lightning events are unknown. 

Although exact locations are not available, we can estimate preliminary statistics from the number of bursts\footnote{\citet[][priv. com.]{hart2014b} defined a burst as an event of at least one second in duration and separated from other events by at least one second.} observed by the \textit{Venus Express}. \citet[][priv. com.]{hart2014b} counted 293 bursts in total with varying duration during three Venus-years (between 2012 and 2013). Obtained flash densities and their possible applicability are shown in Table \ref{table:plan} and discussed in Sect. \ref{sec:exopl}.

\subsection{Giant gas planets} \label{sec:ggp}

Optical and radio observations confirmed the presence of lightning on both giant gas planets, Jupiter and Saturn. Due to the position of the spacecraft, the existing data are limited to specific latitudes and observational times for both planets. Bearing in mind these limitations, i.e. we do not have data from the whole surface of the planet or from continuous observations for a longer period of time (e.g. a year), we use the available data to estimate flash densities for the whole globe of the planet (Table \ref{table:plan}), assuming that at least a similar lightning activity can be expected inside their atmospheres.

\subsubsection{Jupiter} \label{subs:jup}

In 1979 \textit{Voyager 1} and \textit{2} detected lightning flashes on Jupiter \citep{cook1979}. Sferics\footnote{Lightning induced electromagnetic pulses in the low-frequency (LF) range with a power density peak at 10 kHz (for Earth-lightning) \citep{aplin2013}. However, since only radio emission in the higher frequency range can penetrate through a planet's ionosphere, high frequency (HF) radio emission caused by lightning on other planets are also called sferics \citep{desch2002}. Sferics are the result of the electromagnetic field radiated by the electric current flowing in the channel of a lightning discharge \citep{smyth1976}.} were detected inside Jupiter's atmosphere by the \textit{Galileo} probe in 1996 \citep{rinnert1998} and whistlers in the planet's magnetosphere $\sim 20$ years earlier by the \textit{Voyager 1} plasma wave instrument \citep{gurnett1979}. The SSI (Solid State Imager) of the \textit{Galileo} spacecraft observed lightning activity directly on Jupiter during two orbits in 1997 (C10, E11) and one orbit in 1999 (C20). The surveyed area covers more than half of the surface of the planet \citep{little1999}.

\citet{little1999} estimated a lower limit for flash densities on Jupiter to be $4.2 \times 10^{-3}$ flashes km$^{-2}$ year$^{-1}$ based on \textit{Galileo} observations. This value agrees well with the values estimated from the \textit{Voyager} measurements  \citep[$4 \times 10^{-3}$ flashes km$^{-2}$ year$^{-1}$,][]{borucki1982}. \citet{dyudina2004} analysed the same data set and complemented it with \textit{Cassini} observations. In 2007 \textit{New Horizons} observed polar (above $60$\textdegree latitude south and north) lightning on Jupiter with its broadband camera (0.35 - 0.85 $\mu$m bandpass). \textit{New Horizons} found almost identical flash rates for the polar regions on both hemispheres (N: 0.15 flashes s$^{-1}$, S: 0.18 flashes s$^{-1}$).

Correlating lightning flashes with dayside clouds in the \textit{Cassini} data, \citet{dyudina2004} found that lightning occurs on Jupiter in dense, vertically extended clouds that may contain large particles \citep[$\sim 5 \mu$m,][]{dyudina2001, dyudina2004}, typical for terrestrial thunderstorms. However, they also note that lightning observed by \textit{Voyager 2} is not always correlated with these bright clouds, meaning that the low number of small bright clouds does not explain the amount of lightning detected by \textit{Voyager 2}, which observed fainter flashes at higher latitudes than \textit{Cassini} did \citep{borucki1992, dyudina2004}.

\subsubsection{Saturn} \label{subs:sat}

Lightning-induced radio emission on Saturn (Saturn Electrostatic Discharges, SEDs), was first observed by \textit{Voyager 1} during its close approach in 1979 \citep{warwick1981}. The short, strong radio bursts from Saturnian thunderstorms were detected again by the RPWS (Radio and Plasma Wave Science) instrument of the \textit{Cassini} spacecraft in 2004 \citep{fischer2006} and have been recorded since then. SEDs were confirmed to be a signature of lightning activity by the \textit{Cassini} spacecraft. Based on its data, studies associated the radio emission with clouds visible on the images \citep{dyudina2007}. The first Saturnian lightning detection in the visible range (by \textit{Cassini} in 2009) was reported by \citet{dyudina2010}. 

On 17 August 2009 images of Saturn's night side were taken by \textit{Cassini}. Lightning flashes were located on a single spot of the surface at $\sim -36$\textdegree latitude \citep{dyudina2010}. On 30 November 2009 flashes were observed at about the same latitude as before. The flash rate from these observations is 1-2 min$^{-1}$ \citep{dyudina2013}. \citet{dyudina2013} reported further lightning observations on the dayside by \textit{Cassini} at latitude $35$\textdegree north. A new, much stronger storm was observed on 26 February 2011 between latitudes $30$\textdegree $- 35$\textdegree north. A flash rate of 5 s$^{-1}$ was estimated for this storm \citep{dyudina2013}. In the meantime simultaneous SED observations were conducted with the \textit{Cassini}-RPWS instrument between $\sim 2 - 16$ MHz (the first value is the low cutoff frequency of Saturn's ionosphere, while the second one is the instrumental limit). SED rates and flash rates vary for the three storms. Radio (SED) observations were carried out in 2004-2006 by the RPWS instrument. The different storms were observed in different antenna mode, which have different sensitivity. When calculating the SED rates, \citet{fischer2006} took into account the instrument mode as well. The storms and SED episodes are listed in \citet{fischer2006}, their table 1. They found SED rates varying between 30 - 87 h$^{-1}$. Two more SED storms (D and E) were observed in 2005 and 2006 with SED rates much higher than before (367 h$^{-1}$) \citep{fischer2007}. \citet{fischer2011} analysed the SED occurrence of the 2011-storm that started in early December 2010, and found the largest SED rates ever detected on Saturn, to be 10 SED s$^{-1}$. This results in, on average, 36000 SED h$^{-1}$, $\sim 98$ times larger than the SED rate of the largest episode of storm E from 2006.

\subsubsection{Lightning climatology on Jupiter and Saturn} \label{subs:climatjs}

\begin{table*} 
\begin{threeparttable}
 \small 
 \centering
 \begin{minipage}{230mm}
 \caption{Lightning flash densities ($\rho_{\rm flash}$) from four Solar System planets. Exoplanetary examples are also listed under six categories where the flash densities were considered. All values are based on observations. Flash densities are calculated over a year defined in Earth-days, and an hour. Hourly densities are used for estimating lightning activity on exoplanets and brown dwarfs. Yearly $\rho_{\rm flash}$ are calculated in Earth-, Venusian-, Jovian-, or Saturnian-years explained in the text (Sect. \ref{subs:climatjs}).}
  \begin{tabular}{@{}lllllll@{}}	
	\hline 
	Planet & Region & Instrument\tnote{(1)} & \vtop{\hbox{\strut Average yearly $\rho_{\rm flash}$}\hbox{\strut [flashes km$^{-2}$ year$^{-1}$]}} & \vtop{\hbox{\strut Average hourly $\rho_{\rm flash}$}\hbox{\strut [flashes km$^{-2}$ hour$^{-1}$]}} & Exoplanet type & Example \\
	\hline \hline
	\multirow{5}{*}{Earth} & global & LIS/OTD & 2.01 & $2.29\times 10^{-4}$ & Earth-like planet & Kepler-186f \\
	 & continents & LIS-scaled WWLLN & 17.0 & $1.94 \times 10^{-3}$ & \multirow{2}{*}{\vtop{\hbox{\strut Rocky planet with}\hbox{\strut no liquid surface}}} & \multirow{2}{*}{\vtop{\hbox{\strut Kepler-10b}\hbox{\strut 55 Cnc e}}} \\
	 & & LIS/OTD & 28.9 & $3.30 \times 10^{-3}$ & & \\
	 & oceans & LIS/OTD & 0.3 & $3.42 \times 10^{-5}$ & Ocean planet & Kepler-62f \\
	 & & LIS-scaled WWLLN & 0.6 & $6.85 \times 10^{-5}$ & & \\
	\hline
	Venus & global\tnote{(2)} & Venus Express & $2.12\times 10^{-7}$ & $3.64 \times 10^{-11}$ & Venus-like planet & Kepler-69c \\
	\hline
	\multirow{3}{*}{Jupiter} & \multirow{3}{*}{global} & \multirow{3}{*}{\vtop{\hbox{\strut Galileo\tnote{(3)}}\hbox{\strut New Horizons}}} & \multirow{3}{*}{\vtop{\hbox{\strut $2.46 \times 10^{-2}$}\hbox{\strut 0.15}}} & \multirow{3}{*}{\vtop{\hbox{\strut $2.37 \times 10^{-7}$}\hbox{\strut $1.43 \times 10^{-6}$}}} & giant gas planets & \vtop{\hbox{\strut HD 189733b}\hbox{\strut GJ 504b}}\\ \cdashline{6-7}
	 & & & & & brown dwarfs & Luhman-16B \\
	\hline
	\multirow{3}{*}{Saturn} & \multirow{3}{*}{global} & \multirow{3}{*}{\vtop{\hbox{\strut Cassini (2009)}\hbox{\strut Cassini (2010/11)}}} & \multirow{3}{*}{\vtop{\hbox{\strut $1.53 \times 10^{-2}$}\hbox{\strut 1.31}}} & \multirow{3}{*}{\vtop{\hbox{\strut $8.20 \times 10^{-8}$}\hbox{\strut $5.09 \times 10^{-6}$}}} & giant gas planets & \vtop{\hbox{\strut HD 189733b}\hbox{\strut GJ 504b}}\\ \cdashline{6-7}
	 & & & & & brown dwarfs & Luhman-16B \\
	\hline 
  \label{table:plan}
  \end{tabular}
  \begin{tablenotes}
	\item[1] Flash densities are calculated from the data gathered by these instruments
	\item[2] Based on whistler observations, assuming 1 whistler/flash  
	\item[3] Excluding detections during the C20 orbit \citep{gierasch2000, dyudina2004}. 
  \end{tablenotes}
 \end{minipage}
\end{threeparttable}
\end{table*}

\begin{figure}
  \centering
  \includegraphics[width=\columnwidth, trim=0cm 0cm 0cm 0cm]{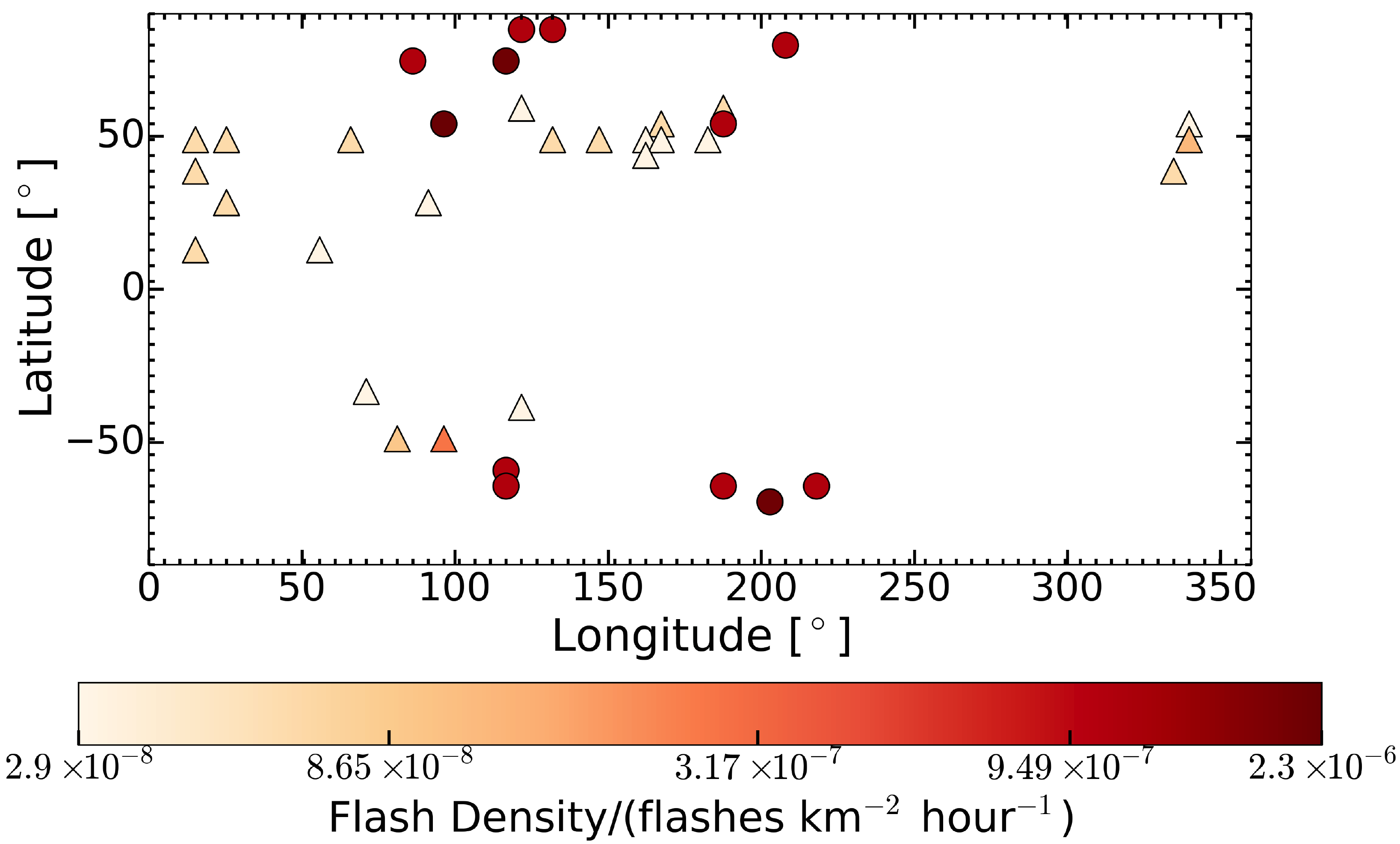}
  \caption{Jupiter lightning occurrence. The colours show the number of flashes averaged in a $5$\textdegree $\times 5$\textdegree area box on the surface of the planet in an hour on a logarithmic scale. Triangles: \textit{Galileo} data \citep[year: 1997,][]{little1999}, Circles: \textit{New Horizons} data \citep[year: 2007,][]{baines2007}. The 10-year gap between the two data sets implies that the plotted lightning flashes are from two different storms.}
  \label{fig:jup}
\end{figure}

\begin{figure}
  \centering
  \includegraphics[width=\columnwidth, trim=0.5cm 0cm 0cm 0cm]{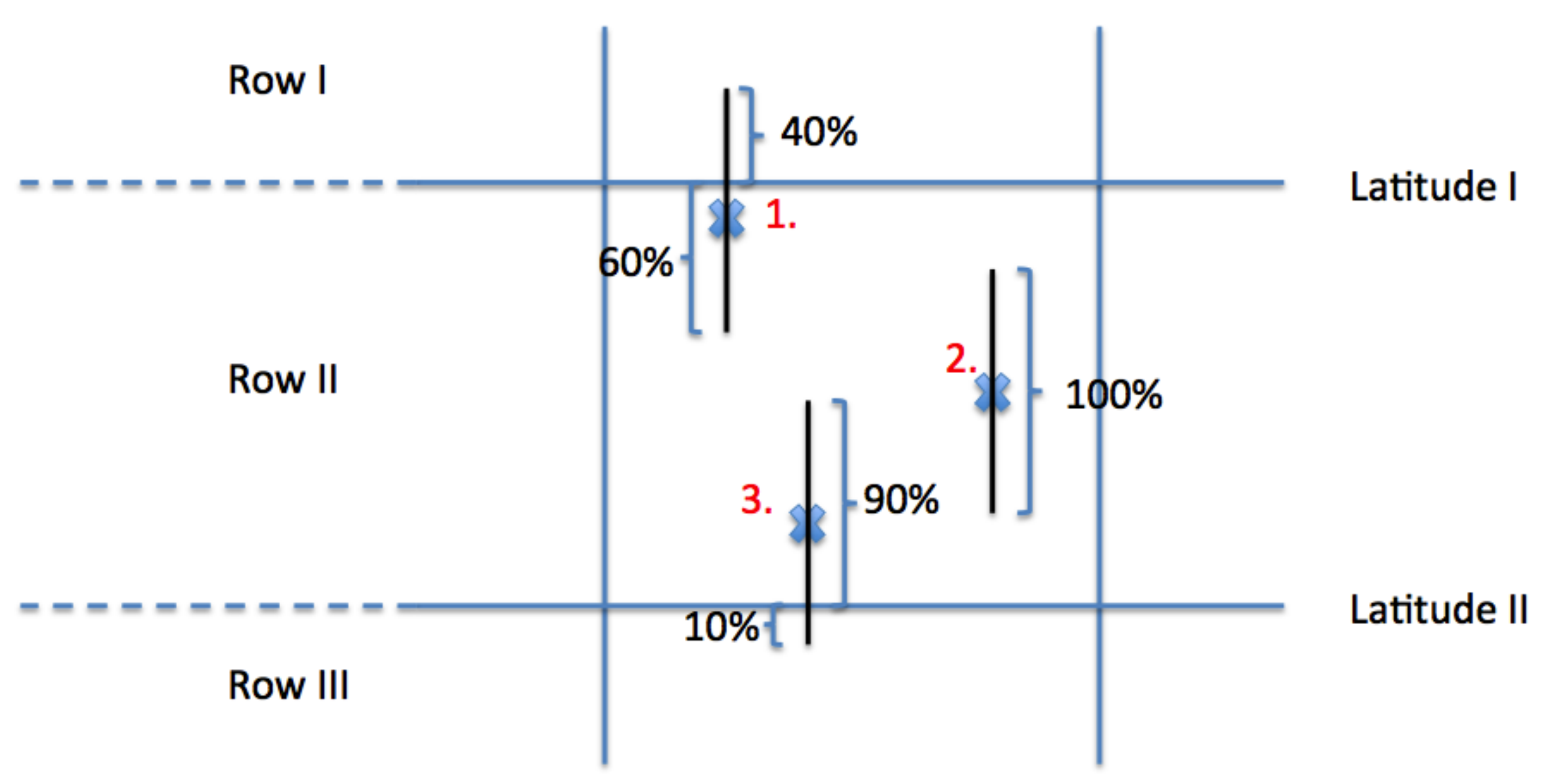}
  \caption{Sketch of grid cells and correction of lightning flash locations. Latitudes I and II define the top and bottom boundaries of a cell. Blue $\times$ signs show the position of the lightning flash with latitude and longitude coordinates. The black error bars are calculated from the spatial resolution of the instrument (Appendix \ref{app:b}). The correction is based on the length of the error bar. When counting the flashes in one grid cell the individual flashes are summed up based on what portion of the full error bar is in the particular cell. E. g. flash 1 is counted 0.6 times in the Row II cell and 0.4 times in the Row I cell; flash 2 is counted as 1 in Row II; flash 3 adds 0.9 times to Row II and 0.1 times to Row III. Adding up we have 0.4 flashes in Row I, $1 + 0.6 + 0.9 = 2.5$ flashes in Row II and 0.1 flashes in Row III, in this example. (The error bars on the figure are for illustration.)}
  \label{fig:grid}
\end{figure}

Data for Jupiter were taken from \citet[][table 1]{little1999}, \citet[][table 1]{dyudina2004} and \citet[][table 1]{baines2007}. \citet{dyudina2004}, their table 1, also lists lightning detections from \textit{Galileo}'s C20 orbit partly based on \citet{gierasch2000}. However, we do not have any information on the occurrence rate of lightning from this orbit, or the coordinates of the observed flashes. Therefore, we did not include these detections in our study. Similarly, no observed coordinates, or flash number estimates are given for the lightning storms observed by \textit{Cassini}, listed in \citet{dyudina2004}, which are also omitted from our study. We summarize these observational data in Figure \ref{fig:jup}, which shows the total number of flashes in an hour (logarithmic scale), averaged in $5$\textdegree $\times 5$\textdegree area boxes over the surface of Jupiter. As explained in Fig. \ref{fig:grid} and Appendix \ref{app:b}, we corrected the spatial (latitudinal) coordinates of the flashes from the \textit{Galileo} data with the pointing error of the instrument calculated from the spatial resolutions given in \citet{little1999}.\footnote{We note that the spatial resolution of the \textit{Galileo} satellite is much finer than the grid set up by us. However, flashes close to the grid edges may overlap two grid cells if the error bars are considered, as described in Fig. \ref{fig:grid}, in which case it is worth applying these error calculations. Our detailed explanation of error calculations can be found in Appendix \ref{app:b}.} The same correction was done for the \textit{New Horizons} data based on spatial resolutions from \citet{baines2007}.

\begin{figure}
  \begin{center}
  \includegraphics[width=\columnwidth, trim=0cm 0cm 0cm 0cm]{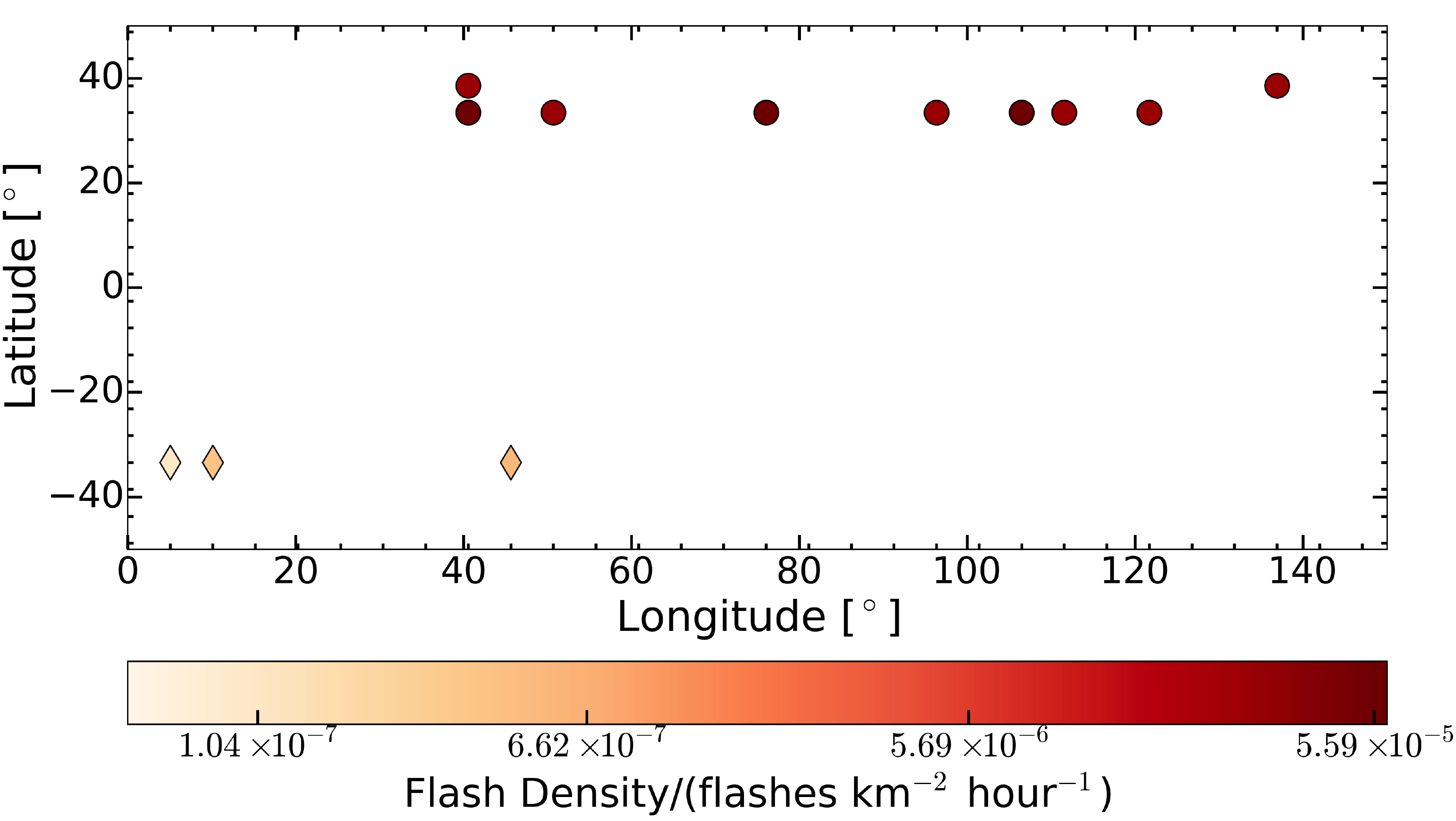}
  \includegraphics[width=\columnwidth, trim=0cm 0cm 0cm 0cm]{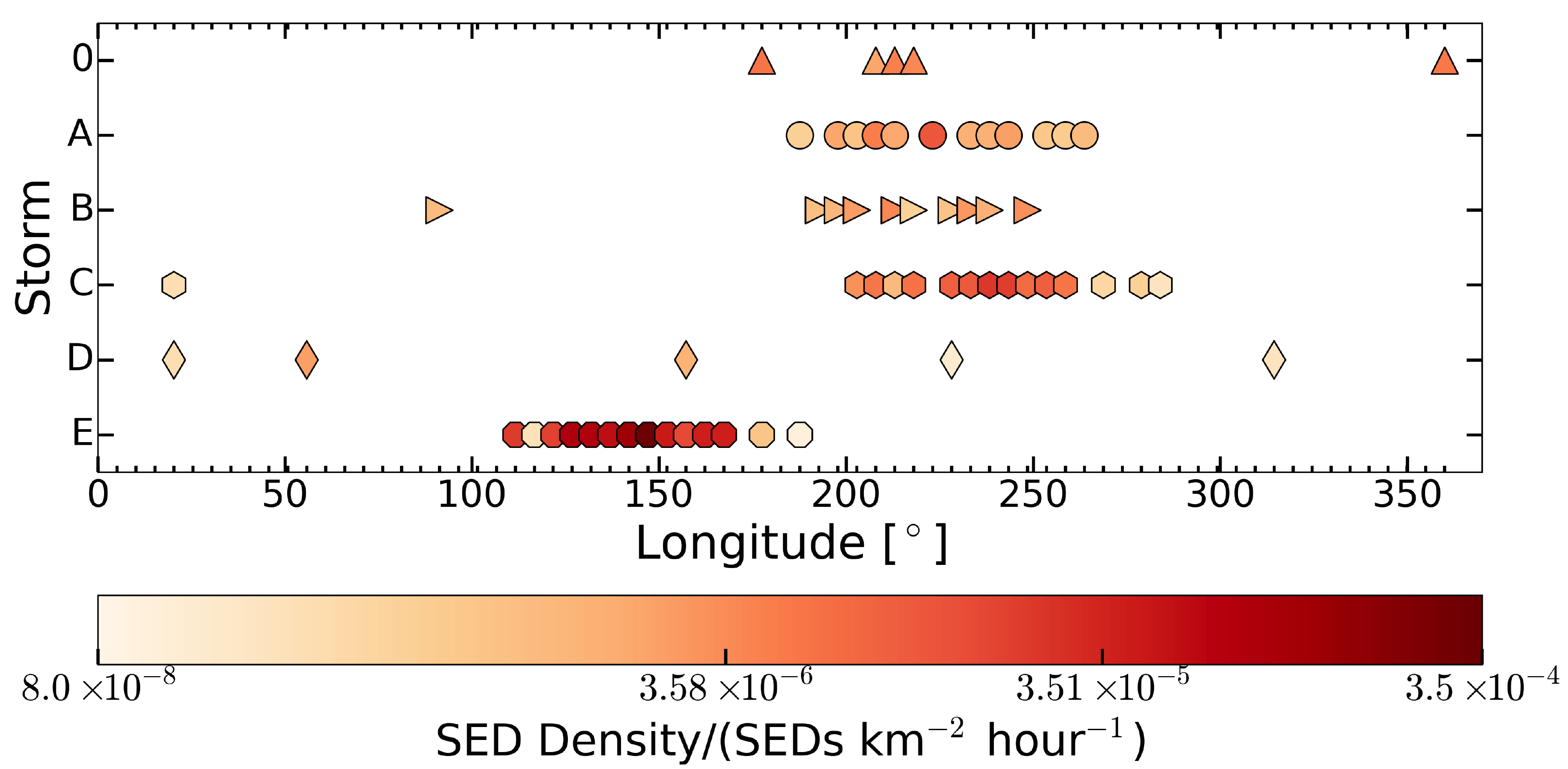}
  \end{center}
  \caption{\textbf{Top}: Saturnian optical lightning occurrence from \textit{Cassini} observations. Data \citep{dyudina2013} are from the years 2009 (diamonds) and 2011 (circles). The data shown are from two different storms. Shown surface region: $\pm45$\textdegree latitude, $0$\textdegree$-150$\textdegree longitude. \textbf{Bottom}: Radio lightning emission, SED, occurrence on Saturn from \textit{Cassini}-RPWS observations in $2004-2007$ \citep{fischer2006, fischer2007}. All SED storms shown on the plot appeared at the $-35$\textdegree latitude region. The figure shows the SED density of each storm plotted against the planetary longitude. The colours on both images show the flash/SED densities averaged in a $5$\textdegree $\times 5$\textdegree area box on a logarithmic scale.
}
  \label{fig:7}
\end{figure}

Saturnian optical data were taken from \citet[][table A1]{dyudina2013}. They list, amongst others, latitudes, longitudes, times of observations, exposure times and spatial resolution. The top panel of Fig. \ref{fig:7} shows the spatial distribution of lightning flashes observed on Saturn in 2009 (diamonds) and 2011 (circles), between latitudes $\pm45$\textdegree  and longitudes $0$\textdegree$-150$\textdegree. The concentration around $\pm 35$\textdegree latitudes is clearly seen. The spatial coordinates of the data were corrected (Appendix \ref{app:b}) with the spatial resolution of the instrument taken from \citet[][Supplement]{dyudina2013}. Similarly to the optical flash observations, we used data taken from \citet{fischer2006} and \citet{fischer2007} for SED measurements. The bottom panel of Fig. \ref{fig:7} shows the SED density on Saturn for 6 different storms, which all appeared on $-35$\textdegree latitude. SED observations were reported from the 2011 storm \citep{dyudina2013}, however, because of the lack of the spatial coordinate information, we do not plot them in Fig. \ref{fig:7}.

The spacecraft observing Jupiter (e.g. \textit{Voyager}, \textit{Galileo}) have found that Jovian lightning activity has a local maximum near $50$\textdegree N \citep[Fig. \ref{fig:jup}; see also][]{little1999}. This might be a consequence of the increasing effect of internal heating compared to solar heating at this latitude. Here, convection is more effective producing thunderclouds with lightning \citep{baines2007}. Solar heating would suppress this effect. \citet{zuchowski2009} modelled the meridional circulation in stratospheric and tropospheric heights of Jupiter's atmosphere, and found an upwelling in the zones and downwelling in the belts in stratospheric levels. However, at lower atmospheric heights upwelling was found in the belts, which allows the formation of water clouds and lightning discharges, just like observations indicate \citep{little1999, ingersoll2000, zuchowski2009}. \citet{dyudina2013} found that on Saturn lightning occurs in the diagonal gaps between large anticyclones. These gaps are similar to Jovian belts, composed of upwelling, convective thunderstorms \citep{dyudina2013, read2011} (Fig. \ref{fig:7}). We do not attempt to compare lightning occurrence via longitudes, since due to the drift of the storms that would not be a valid approach without correcting for this drift.

The results in Table \ref{table:plan} include hourly and yearly average flash densities obtained for the Solar System planets.\footnote{The results in Table \ref{table:plan} are based on positive detections of lightning. This is important especially on Saturn, where most of the time no storm was observed resulting in 0 flash densities \citep{fischer2011b}.} Yearly flash densities were calculated for a year defined in Earth-days (24-hour days), and they represent the length of a year on the appropriate planet. For example: when calculating flash rates (flashes year$^{-1}$) for Jupiter, we used a Jovian-year of 4330 days and not 365-Earth days (apart from Earth lightning flash rates). Similarly we define Venusian- and Saturnian-years too. Global flash densities were estimated for all of the planets (Table \ref{table:plan}). For Earth we distinguish between continental and oceanic rates. The values in Table \ref{table:plan} for the two latter regions are calculated from LIS/OTD (larger value) and LIS-scaled WWLLN (lower value) data. Similarly, the larger values for Jupiter are estimated from  \textit{New Horizons} data, while lower ones are based on \textit{Galileo} data. For Saturn, the larger values are based on data from the giant storm in 2011, while the lower ones are from the 2009-storm.

We calculated flash rates (flashes year$^{-1}$ or flashes hour$^{-1}$; $R_{\rm flash}$) for Jupiter and Saturn for each of the images taking into account the exposure times as given by:

\begin{equation} \label{eq:1}
R_{{\rm flash},i} = \frac{n_i}{t_{{\rm exp},i}} C,
\end{equation}
\noindent where $n$ is the number of flashes detected in image $i$, $t_{\rm exp}$ is the exposure time of the image in seconds, and $C$ is a unitless scaling factor, which converts the time units from seconds to hours or years.\footnote{We do not analyse flash rates. For more details about flash rates see \citet{dyudina2013}, their table 2.} The flash density (flashes unit-time$^{-1}$ km$^{-2}$, $\rho_{\rm flash}$), is calculated from Eq. (\ref{eq:2}), with $R_{\rm flash}$, given by Eq. (\ref{eq:1}).

\begin{equation} \label{eq:2}
	\rho_{\rm flash} = \frac{\sum_{i=1}^{i=N}{R_{{\rm flash},i}}}{A_{\rm surv}},
\end{equation}
\noindent where $N$ is the total number of images and $A_{\rm surv}$ is the total surveyed area: $A_{\rm surv}^{\rm Galileo} = 39.5 \times 10^9$ km$^2$ \citep{little1999}, $A_{\rm surv}^{\rm New Horizons} = 8.0 \times 10^9$ km$^2$ \citep{baines2007}\footnote{Calculated based on \citet{baines2007}, information on image resolution in footnote 15 and surveyed latitude range in figure 1.}. $A_{\rm surv}$ for the 2009 storm on Saturn is the 30\% of Saturn's surface area \citep[][Supplement]{dyudina2010}, and $A_{\rm surv}$ for the 2011 storm is the total area of Saturn based on the fact that the \textit{RPWS} instrument detected only one SED storm on the whole planet at a time \citep{fischer2006, fischer2007}.

The flash densities for Jupiter derived here are different from previously published values \citep[$\sim 4 \times 10^{-3}$ flashes km$^{-2}$ year$^{-1}$,][]{little1999, borucki1982}, which is the result of converting exposure times, which are given in seconds, to years. For example, from the \textit{Galileo} data we obtain a flash density of 0.02 flashes km$^{-2}$ year$^{-1}$ when we take the length of a Jovian year to be the number of days Jupiter orbits the Sun, 4330 days. This way we get a flash density an order of magnitude higher than previously estimated \citep[e.g.][]{little1999}. However, when we determine the flash rate (flashes year$^{-1}$) considering a year to be 365 days long, the way it is done in \citet{little1999}, and divide it by the \textit{Galileo} survey area, our result becomes the same order of magnitude but twice lower than the one in \citet{little1999}, or $2 \times 10^{-3}$ flashes km$^{-2}$ year$^{-1}$ compared to $4 \times 10^{-3}$ flashes km$^{-2}$ year$^{-1}$. This factor of two is a reasonable difference, since we do not consider over-lapping flashes in our work (U. Dyudina, private communication). \citet{little1999} calculated flash densities saying that on average there were 12 flashes detected in one storm. They multiplied this by the number of storms observed (26, their table I) and divided by an exposure time of 59.8 s and the total survey area of $39.5 \times 10^9$ km$^2$. In our approach, we took the data from table I in \citet{little1999} and table 1 of \citet{dyudina2004}, counted the flashes on each frame, assuming that one "lightning spot" in table 1 of \citet{dyudina2004} corresponds to one lightning flash, then divided that number with the exposure time (in years or hours, with 1 year on Jupiter being $3.73 \times 10^8$ s) of the frame. After summing up these flash rates, we divided the result with the total surveyed area of $39.5 \times 10^9$ km$^2$. Therefore, the differences between previously calculated flash densities and flash densities listed in Table \ref{table:plan} are the result of converting exposure times to years. However, for our purposes we only use hourly flash densities, which do not depend on the length of a year.

The above derived formulas and the resulting values listed in Table \ref{table:plan} involve various uncertainties, which also affect the comparability. The flash rate, $R_{\rm flash}$, depends on the number of detected flashes ($n_i$) at a certain time determined by the exposure time ($t_{{\rm exp},i}$). $n_i$ is affected by instrumental sensitivity, the time of the survey (seasonal effects on lightning occurrence) and the place of the survey (different lightning occurrence over different latitudes and surface types, Figs. \ref{fig:1}-\ref{fig:4}). The flash density, $\rho_{\rm flash}$, is derived from $R_{\rm flash}$ (Eq. \ref{eq:2}). Uncertainties also rise from the not-precise determination of total surveyed area. Baring in mind these limitations of the data and uncertainties in the values in Table \ref{table:plan}, we apply our results of flash densities on exoplanets and brown dwarfs in Sect. \ref{sec:exopl}.

\subsection{Energy distribution} \label{subs:endis}

\begin{figure}
  \centering
  \includegraphics[width=\columnwidth, trim=0cm 0cm 0cm 0cm]{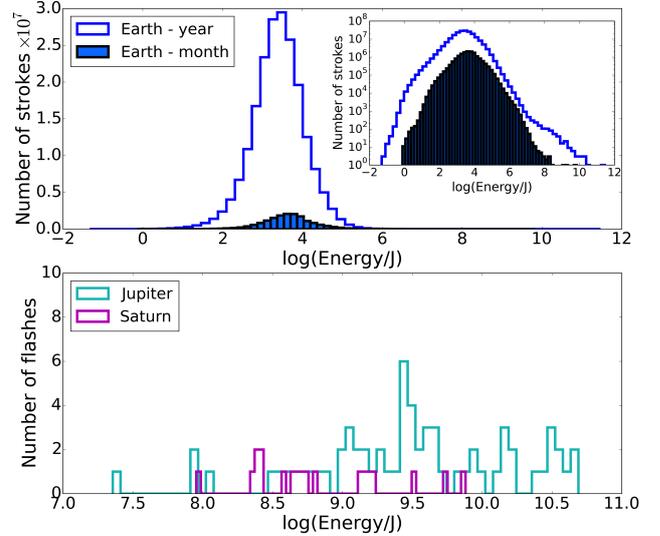}
  \caption{Radio energy distribution of lightning strokes from Earth (WWLLN, 2013; top) based on data from the whole year compared to a month (2013, December), which do not distinguish between IC and CG lightning, and optical energy distribution (calculated from measured powers, see Sect. \ref{subs:endis}) of Jovian and Saturnian flashes (bottom). The data from Jupiter (\textit{Galileo}, 1997; \textit{New Horizons}, 2007) and Saturn (\textit{Cassini}, 2009, 2011) are both from less than an hour of observations (about 50 minutes for Jupiter and 10 minutes for Saturn \citep[exposure times in][]{little1999, baines2007, dyudina2013}). The number-energy distributions of the Earth-data seems to be self-similar in time as it has the same shape if plotted for a month or for a year. The inset plot (top panel) shows a log-log scale of the Earth data.
}
  \label{fig:endist}
\end{figure}

Figure \ref{fig:endist} summarizes the number distribution of stroke energies for Earth (top), and number distribution of flash energies for Jupiter and Saturn (bottom). For Earth we used WWLLN data from 2013, while for the outer planets we included all data from \textit{Galileo}, \textit{New Horizons} and \textit{Cassini}. \citet{dyudina2004} lists the power [$W = {\rm J} {\rm s}^{-1}$] of lightning as observed by the \textit{Galileo} probe (their table 1, column 11). Following the procedure in \citet[][eq. 1]{dyudina2013} where they treated storms as continuously flashing steady light sources and each flash as a patch of light on a Lambertian surface, we converted the measured power values to energies by multiplying them with the exposure time. On Earth most of the strokes have radio energies of the order of $10^3-10^{3.5}$ J. This indicates that less energetic lightning flashes, due to their large number, are likely to be more significant for chemically changing the local gas in large atmospheric volumes. However, a detailed modelling of the structure and size of discharge channels are required for drawing more definite conclusions.

We need to be careful with overinterpretation of the directly accessible data; however, the knowledge gained about their limitations is useful when discussing lightning observability. Due to instrumental limitations (detection threshold), only the most energetic lightning events are detectable. This is particularly prominent in the Saturnian and Jovian data (Fig. \ref{fig:endist}, bottom panel). It seems impossible to find the peak of the energy distribution, being lower than the detection limit, on Saturn and Jupiter just by extrapolating the limited number of data points. However, we may assume that most of the lightning flashes will cluster around one energy also for Jupiter and Saturn, and that this peak in flash numbers will move to higher energies compared to Earth. This expectation is based on the fact that the underlying physics (i.e. electron avalanches develop into streamers in an electric potential gradient)  is only marginally affected by the chemical composition of the atmospheric gas \citep[e.g.][]{helling2013}, and the fact that Jupiter's and Saturn's clouds have a larger geometrical extension and, hence, a larger potential difference than on Earth. \citet{bailey2014} showed that a larger surface gravity, like on Jupiter compared to Earth, leads to larger geometrical extension of a discharge event with higher total dissipation energies. \citet{dyudina2004} suggest that their lightning power values derived from observations are underestimates, as $25\%$ of the lightning spots are saturated in the \textit{Galileo} images, they do not consider the scattered light on clouds, which may dim the flashes by a couple of orders of magnitude \citep{dyudina2002}. This suggests that the observed energies on Jupiter are most likely exceeding the largest lightning energies observed on Earth. From this, one may assume that the peak of the energy distribution of lightning flashes on the gas giant planets also shifts to higher energies. \citet{dyudina2004} analysed the power distribution of optical lightning flashes on Jupiter, considering only flashes recorded by \textit{Galileo}'s clear filter\footnote{385 - 935 nm \citep{little1999}.} (their fig. 7). They showed that the number of flashes with high power is small, which is similar to observations for Earth (similarly: Fig. \ref{fig:endist}, top panel). However, observations result in low detected flash numbers. Moreover, lightning observations in the Solar System have biases towards higher energy lightning. Therefore, \citet{dyudina2004} concluded that lightning frequencies at different power levels cannot be predicted unequivocally.

We also note that \citet{farrell2007} suggested that Saturnian discharges might not be as energetic as they were thought to be ($\sim 10^{12}$ J). They assumed a shorter discharge duration, which would result in lower discharge energies. Their study shows the importance of exploring the parameter space that affects lightning discharge energies and radiated power densities, in order to interpret possible observations of not yet fully explored planets.

\section{Discussing lightning on exoplanets and brown dwarfs} \label{sec:exopl}

\begin{figure}
  \centering
  \includegraphics[width=\columnwidth, trim=0cm 0cm 0cm 0cm]{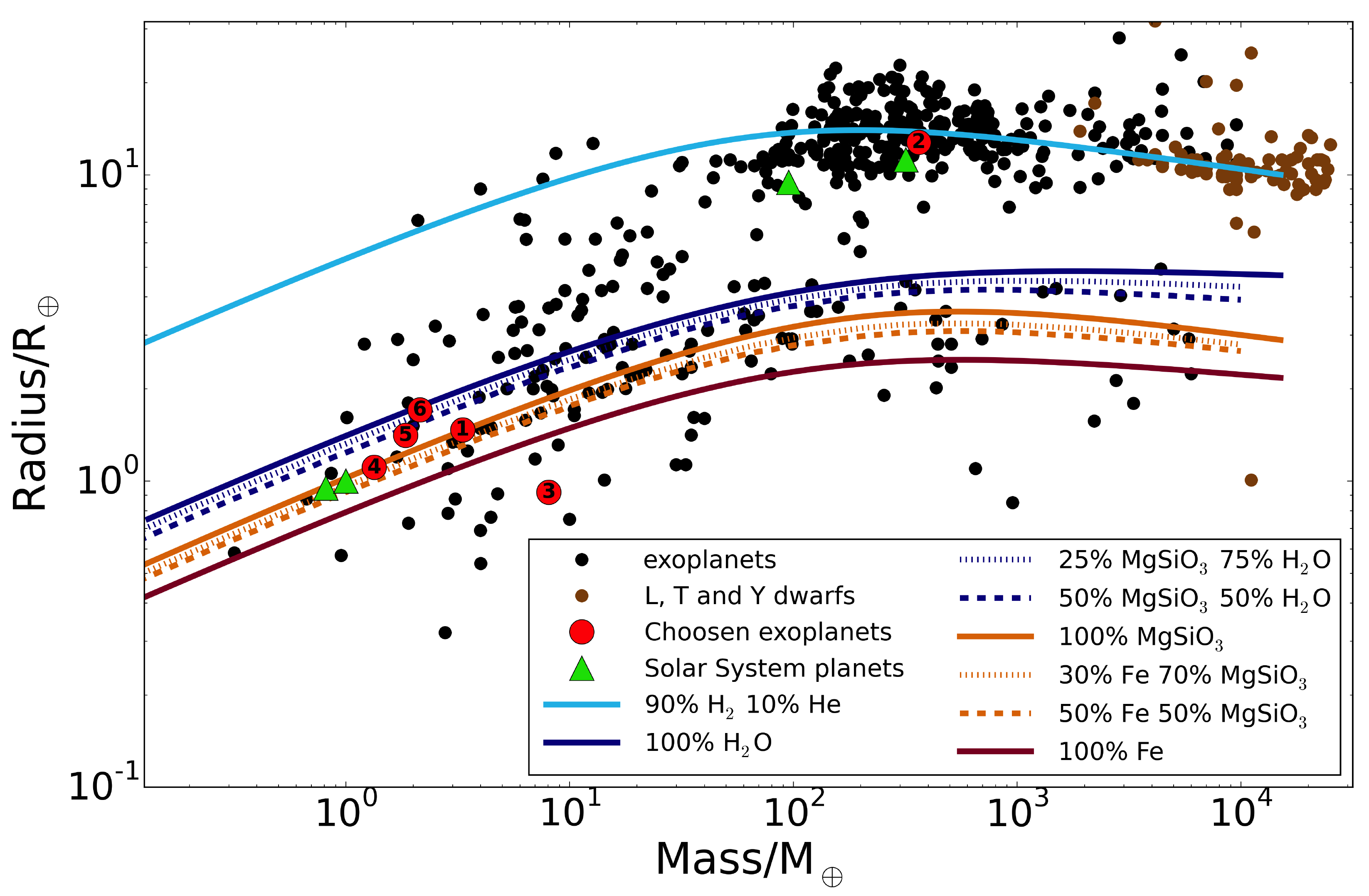}
  \includegraphics[width=\columnwidth, trim=0cm 0cm 0cm 0cm]{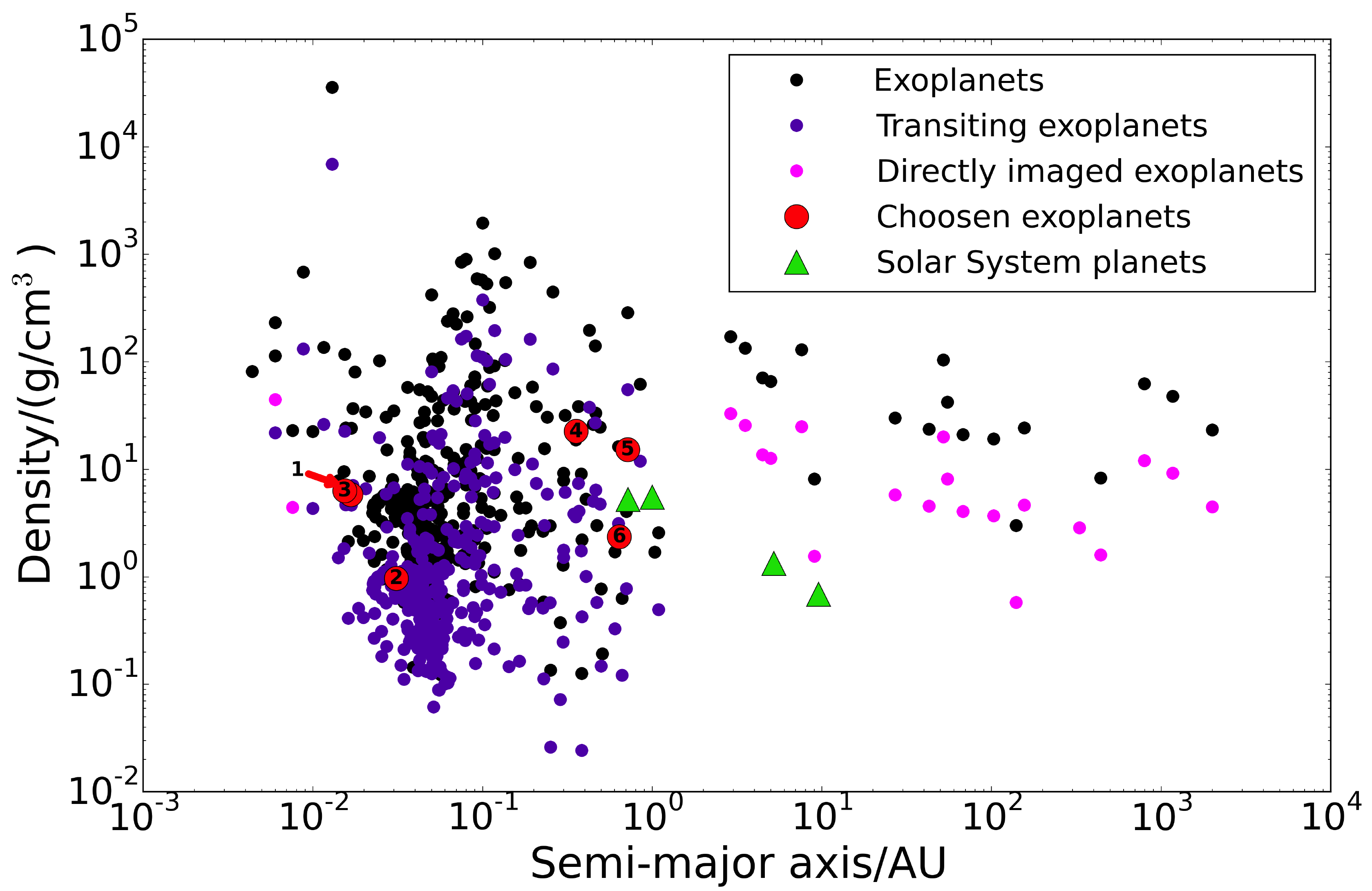}
  \caption{Diversity of known exoplanets and brown dwarfs. Red circles with numbers represent the exoplanet examples used in our case studies (Sect. \ref{sec:casest}): 1 - Kepler-10b, 2 - HD 189733b, 3 - 55 Cnc e, 4 - Kepler-186f, 5 - Kepler-62f, 6 - Kepler-69c. GJ 504b is not on the plots, since no radius is available for this planet. Green triangles indicate Venus, Earth, Jupiter and Saturn.
\textbf{Top}: Relation between mass and radius (in Earth-values: M$_{\oplus}$, R$_{\oplus}$). Black and brown dots represent exoplanets and brown dwarfs, respectively. The lines show mass-radius relationships for various bulk compositions. We note that for some cases the uncertainties in mass and radius are large enough to move the planet from one compositional region to the other. The uncertainties are especially large for Kepler-62f, for which only the upper mass limit is known (Table \ref{table:planet}). However, \citet{kaltenegger2013} estimated the mass of Kepler-62f to be, on average, $\sim 1.85$ M$_{\oplus}$, and we use this value on our figure.
\textbf{Bottom}: Average density ($\rho_{\rm bulk}$ [g/cm$^3$]) vs semi-major axis ($a$ [AU]) of exoplanets. Blue dots indicate transiting planets, magenta dots show directly imaged planets. The density of Kepler-186f, Kepler-62f (see Sect. \ref{sec:exopl}) are mean densities calculated from the radius and mass, while the density of Kepler-10b \citep{dumusque2014}, 55 Cnc e \citep{demory2015} and HD 189733 b \citep{torres2008} are from Markov Chain Monte Carlo analysis of photometric data.}
  \label{fig:mr}
\end{figure}

The Solar System planets, especially Earth, have been guiding exoplanetary research for a long time. Models have been inspired, for example, for cloud formation \citep[e.g.][]{lunine1986, ackerman2001, helling2008, kitzmann2010} and global atmospheric circulation \citep[e.g.][]{dobbs-dixon2013, mayne2014, zhang2014}, and have been used for predictions that reach far beyond the Solar System. Habitability studies \citep[e.g.][]{kaltenegger2007, betremieux2013} have been conducted based on signatures, called biomarkers \citep{kaltenegger2002}, appearing in Earth's spectra.

In this paper we use lightning climatology studies from Solar System planets for a first discussion on the implications of potential lightning occurrence on exoplanets and brown dwarfs. Though the data discussed in Sections \ref{sec:earth} and \ref{sec:ggp} are limited to radio and optical observations, these studies are also useful for the better understanding of the evolution of extrasolar atmospheres through, for example, changes in the chemistry as a result of lightning discharges \citep{rimmer2016}.

Figure \ref{fig:mr} shows the diversity of extrasolar planetary objects with respect to their mean composition (top panel) and their distance from the host star (bottom panel). Figure \ref{fig:mr} also includes the Solar System planets discussed in this paper and the exoplanets considered in the next section (green triangles and red circles, respectively). We include L, T and Y brown dwarfs, for which the masses and radii were taken from a brown dwarf list.\footnote{johnstonsarchive.net/astro/browndwarflist.html - by Wm. Robert Johnston. Several brown dwarf lists can be found on the internet, though most of them do not include size and mass parameters. A well-composed, continuously updated list of brown dwarfs can be found on https://jgagneastro.wordpress.com/list-of-ultracool-dwarfs/ by J. Gagne, where coordinates, identifiers, proper motions, etc. are listed, however no radius and mass information are added.}

The top panel of Fig. \ref{fig:mr} includes density curves for different bulk compositions, including pure water, iron and enstatit (MgSiO$_3$) and the mix of these. We also include the line for a 90\% H$_2$ 10\% He composition.\footnote{These lines were calculated by solving the equations for hydrostatic equilibrium and the mass of a spherical shell. For all compositions except H$_2$/He, we assume a modified polytrope for the equation of state, $\rho = \rho_0 + cP^n$ with the parameters ($\rho_0,c,n$) taken from \citet{seager2007}. For H$_2$/He, we use the equation of state from \citet{militzer2013}.} The density lines visualize the diversity of the global chemical composition of extrasolar bodies. The gas giants and brown dwarfs line up around the H$_2$/He line (light blue line), possible water words and Neptune-like planets follow the lines with H$_2$O content (dark blue lines), while rocky planets, super-Earths are found around the MgSiO$_3$ composition lines (orange lines). A populated region above the pure H line includes the inflated hot Jupiters, whose radii are larger due to the close vicinity to the host star (see Fig. \ref{fig:mr}, top panel). Figure \ref{fig:mr} (bottom) further illustrates that many of the presently confirmed exoplanets reside considerably closer to their host star than any of the Solar System planets. Therefore, the characteristics of the host star will also be of interest for our purpose of discussing potential candidates for further theoretical and observational lightning studies.

The diversity of observed extrasolar planets implies a large variety of atmospheric chemistry and dynamics. Some planets will have atmospheric chemical compositions similar to brown dwarfs, others will be more water or methane dominated and therefore maybe more comparable to the Solar System planets. The basic physical processes that lead to the formation of clouds (nucleation, bulk growth/evaporation, gravitational settling, element depletion) will be the same, independent of the local chemistry, though their efficiency might differ \citep[e.g.][]{helling2014b}. According to transit spectrum observation, extrasolar planets form clouds in their atmosphere \citep[e.g.][]{sing2009, sing2013, sing2015}, and \textit{Hubble Space Telescope} and \textit{Spitzer} observations have suggested that these atmospheres are very dynamic \citep[e.g.][]{knutson2008, knutson2012, buenzli2014, buenzli2015}. The study of possible cloud particle ionization has only begun in the context of extrasolar planets and brown dwarfs \citep[e.g.][]{helling2011b, rimmer2013}. \citet{helling2011b} have shown that triboelectric charging of cloud particles is likely to occur also in extrasolar planetary clouds. Further cloud particle charging will result from cosmic ray impact \citep{rimmer2013} and chemical surface reactions. \citet{helling2013} have demonstrated, based on data by \citet{sentman2004}, that the electric field breakdown, which initializes a lightning discharge does not very strongly depend on the chemical composition of the gas (e.g. their fig. 5). We, therefore, suggest that the Solar System lightning statistics presented here can be used as a first guidance for lightning occurrence on extrasolar planets and brown dwarfs. We, however, note that the Solar System flash rates and densities carry uncertainties as discussed in Sect. \ref{subs:climatjs}.

In order to apply the results of the previous sections on lightning climatology, we group the extrasolar planetary objects into several categories (Sect. \ref{sec:casest}). Bearing in mind the diversity of exoplanets, we choose specific examples for each category, which are discussed in more details to demonstrate why they might be suitable candidates for lightning activity. Figure \ref{fig:mr} shows where these planets (red circles) lie in the (M$_p$, R$_{\rm p}$)-plane and in the (a, $\rho_{\rm bulk}$)-plane compared to the whole ensemble of known exoplanets and brown dwarfs. Section \ref{sec:flashdens} presents the flash densities estimated for the extrasolar category examples. Section \ref{sec:stelact} discusses the challenges arising from the stellar activity of the host stars of planets, and also how this activity may favour the production of lightning on planets. We, however, note that more fundamental modelling of the 3D cloud forming, radiative  atmosphere structure like in \citet{lee2015} and \citet{helling2016}, possibly in combination with kinetic gas-phase modelling like in \citet{rimmer2016} is required to provide quantitative results. In the following, we make a first qualitative attempt of selecting possible candidates for future studies.

\subsection{Case-study categories} \label{sec:casest}

\begin{table*} 
\begin{threeparttable}
 \small 
 \centering
 \caption{Properties of exoplanets and the brown dwarf, Luhman 16B, listed in this paper as examples for further study of lightning activity.}
  \begin{tabular}{@{}lllllll@{}}	
	\hline
	\vtop{\hbox{\strut super-Earth}\hbox{\strut size planet}} & \vtop{\hbox{\strut Mass}\hbox{\strut (M$_p$/M$_\oplus$)}} & \vtop{\hbox{\strut Radius}\hbox{\strut (R$_{\rm p}$/R$_\oplus$)}} & \vtop{\hbox{\strut Density}\hbox{\strut ($\rho_p$/(g cm$^{-3}$))}} & \vtop{\hbox{\strut Semi-major axis}\hbox{\strut (a/AU)}} & \vtop{\hbox{\strut Calculated temperature\tnote{(1)}}\hbox{\strut (T$_{\rm cal}$/K)}} & Reference \\
	\hline
	Kepler-186 f & $0.31-3.77$ & $1.11\substack{+0.14 \\ -0.13}$ & & $0.356 \pm 0.048$ & & \citet{quintana2014} \\
	Kepler-62 f & $ < 35$ & $1.41 \pm 0.07$ & & $0.718 \pm 0.007 $ & $208 \pm 11$ (T$_{\rm eq}$) & \citet{borucki2013} \\
	Kepler-10 b & $3.33 \pm 0.49$ & $1.47\substack{+0.03 \\ -0.02}$ & $5.8 \pm 0.8$ & $0.01685 \pm 0.00013$ & $2169\substack{+96 \\ -44}$ (T$_{\rm eq}$) & \citet{dumusque2014} \\
	55 Cnc e & $8.08 \pm 0.31$ & $1.92 \pm 0.08$ & $6.3\substack{+0.8 \\ -0.7}$ & $0.01544 \pm 0.00009$ & $\sim 2400$ (T$_{\rm eq}$) & \citet{demory2015} \\
	Kepler-69 c & 2.14 & $1.71\substack{+0.34 \\ -0.23}$ & 2.36 & $0.64\substack{+0.15 \\ -0.11}$ & $299\substack{+19 \\ -20}$ (T$_{\rm eq}$) & \vtop{\hbox{\strut \citet{barclay2013}}\hbox{\strut \citet{kane2013}}} \\
	\hline
	\vtop{\hbox{\strut Jupiter}\hbox{\strut size planet /}\hbox{\strut brown dwarf}} & \vtop{\hbox{\strut Mass}\hbox{\strut (M$_p$/M$_{\rm Jup}$)}} & \vtop{\hbox{\strut Radius}\hbox{\strut (R$_{\rm p}$/R$_{\rm Jup}$)}} & \vtop{\hbox{\strut Density}\hbox{\strut ($\rho_p$/(g cm$^{-3}$))}} & \vtop{\hbox{\strut Semi-major axis}\hbox{\strut (a/AU)}} & \vtop{\hbox{\strut Calculated temperature\tnote{(1)}}\hbox{\strut (T$_{\rm cal}$/K)}} & Reference \\
	\hline
	HD 189733 b & $1.14 \pm 0.06$ & $1.14 \pm 0.03$ &  $0.75 \pm 0.08$ & $0.0309 \pm 0.0006$ & $1201 \pm 13$ (T$_{\rm eq}$) & \citet{torres2008} \\
	GJ 504 b & $4.0\substack{+4.5 \\ -1.0}$ & - & - & 43.5 & $510\substack{+30 \\ -20}$ (T$_{\rm eff}$) & \citet{kuzuhara2013} \\
	Luhman 16B & $20-65$ & - & - & - & $1280 \pm 75$K (T$_{\rm{eff}}$) & \citet{faherty2014} \\
	\hline
  \label{table:planet}
  \end{tabular}
  \begin{tablenotes}
	\item[1] T$_{\rm eff}$: effective temperature; T$_{\rm eq}$: equilibrium temperature     
  \end{tablenotes}
\end{threeparttable}
\end{table*}

Transiting planets like Kepler-186f, Kepler-62f, Kepler-10b, 55 Cancri e, Kepler-69c and HD 189733b, directly imaged planets such as GJ 504b, and brown dwarfs like Luhman 16B, are some of the best candidates for the detection of lightning or its effects on the atmosphere.

The spectrum of a transiting exoplanet may contain various information on the atmosphere of the planet, possibly including signatures of lightning. These signatures may be emission or absorption lines either caused by lightning or by non-equilibrium species as a result of lightning activity \citep[e.g.][]{barnun1985,krasnopolsky2006,kovacs2010,bailey2014}\footnote{Lightning may occur anytime throughout a planet's orbit, and its signatures could appear in any observational technique good enough to pick them up. However, currently transiting exoplanets offer the largest numbers of detected exoplanets with techniques related to transit- or occultation-observations being one of the most successful ones in characterizing these objects.}. Directly imaged planets are another category of good candidates for lightning-hunting. They are far enough from their host star, so that the stellar light can be blocked by coronagraphs and the planet's disc can be observed directly. These planets, being far from stellar effects, are comparable to non-irradiated brown dwarfs \citep[e.g.][]{kuzuhara2013, janson2013}. Brown dwarfs are much closer to us than any exoplanet and, in most of the cases, no host star will outshine their signal. Therefore, brown dwarfs are among the best candidates from the sample of objects that we have available (see Fig. \ref{fig:mr}) to detect lightning in their spectrum (e.g. radio, or other suitable means).

Lightning may be an indicator of potentially habitable environments, since it may be essential for the formation of prebiotic molecules and because it carries information about cloud dynamics. Some of the planets that we examine below are suggested to reside in the Habitable Zone (HZ) of their host star. The HZ is usually defined as the region where the incident flux of the star is enough for liquid water to be maintained on the surface of a planet with adequate atmospheric pressure \citep[e.g.][]{kasting1993, kasting2014, kopparapu2013, kopparapu2014}. Habitability is a very hot topic of exoplanetary research, resulting in various studies and concepts of the HZ. Some researchers apply the "water loss" and "maximum greenhouse" limits \citep[e.g.][]{kasting1993,kopparapu2013}, others define the boundaries between "recent Venus" and "early Mars" limits \citep[e.g.][]{kasting2014}, and some use an even more extended HZ concept \citep[e.g.][]{seager2013}. These various HZ definitions show the uncertainty in the precise definition of a habitable planet, which allows us to develop a wider concept of planets with lightning. 

Below, we define six categories guided by the availability of lightning observations from the Solar System planets. These categories are not exclusive but should rather be understood as guides to existing knowledge from the Solar System. We use lightning climatology results from Sections \ref{sec:earth} and \ref{sec:solsys} in order to provide a first estimate of potential lightning occurrence on extrasolar planetary bodies. The objects listed under each category are examples of a larger number of planets/brown dwarfs as demonstrated in Fig. \ref{fig:mr}. The chosen examples have been observed with different techniques before. The properties of the planets considered below and the properties of their host stars are summarized in Tables \ref{table:planet} and \ref{table:star}. 

\begin{itemize}
\item \textbf{Earth-like planets}: Planets with similar continent-ocean fraction as Earth. Studies have shown that, in principle, it is possible to estimate the ocean-land ratio of the surface of the planet by detecting diurnal variability in the photometric light curve of the planet \citep[e.g.][]{ford2001, kawahara2010}. \citet{ford2001} developed a model, which considers Earth as an exoplanet and analysed its light curve with and without clouds. They found significant, potentially detectable, changes in the light curve as the different surfaces (ocean, land, desert) rotated into the view. \citet{kawahara2010} developed a method to reconstruct the surface of a planet using variations in its scattered-light curve. This model was shown to work for an Earth-like surface, however, several assumptions were made, such as cloudlessness or lack of atmospheric absorption. \citet{kawahara2011} used simulated exoplanet light curves from Earth observations by the \textit{EPOXI} mission and demonstrated that the inversion of the light curves recovers the cloud coverage of the planet. By subtracting the cloud features they also showed that the residual maps created from the data trace the continental distribution of Earth. Knowing the ratio of continent-ocean coverage of an exoplanet would help to estimate the lightning occurrence on such planets, however, based on above mentioned studies, it seems retrieving land-ocean fractions on planets needs improvement in observational instrumentation. Regardless, once the tools are available, either continent-ocean surface mapping of a planet may help lightning detections or, vice versa, lightning signal distribution may help the surface mapping of an extrasolar object. We choose our candidate planet for this category based on previous studies. We used the global average flash density from Earth for these planets. 

\begin{itemize}
\item[$\ast$] Example: \textit{Kepler-186f} \citep[][and number 4 in Fig. \ref{fig:mr}]{quintana2014}.
\end{itemize}

The Kepler-186 planetary system is composed of five planets, all with sizes smaller than $1.5$ R$_\oplus$ (Earth radius) \citep{quintana2014}. \citet{quintana2014} reported the discovery of Kepler-186f, the only planet of the five in the system lying in the HZ of the host star. According to their modelling the mass of Kepler-186f can range from $0.31$ M$_\oplus$ (M$_\oplus$: Earth mass) to $3.77$ M$_\oplus$ depending on the bulk composition (from pure water/ice to pure iron composition). In case of an Earth-like composition its mass would be $1.44$ M$_\oplus$. \citet{torres2015} found that Kepler-186f has a 98.4\% chance of being in the HZ of the host star. \citet{bolmont2014} found that with modest amount of CO$_2$ and N$_2$ in its atmosphere, the surface temperature can rise above 273 K and the surface of the planet could maintain liquid water permanently. If Kepler-186f indeed has an Earth-like composition as Fig. \ref{fig:mr} suggests, it may host an atmospheric circulation and convectively active clouds just as Earth, which makes it an interesting candidate of hosting lightning activity.

\item \textbf{Water worlds (Ocean planets)}: Planets with surfaces fully covered by water or very small continent-to-water ocean fractions. The irradiation from the host star can drive strong winds, which may cause the formation of intermittent clouds. Lightning flash density over the Pacific Ocean is used in this analysis. 

\begin{itemize}
\item[$\ast$] Example: \textit{Kepler-62f} \citep[][and number 5 on Fig. \ref{fig:mr}]{borucki2013}.
\end{itemize}

Using Ca H\&K emission index, \citet{borucki2013} concluded that Kepler-62, a K-type main-sequence star, is inactive. Kepler-62f is the outermost planet in the 5-planet system. By calculating the incident flux, \citet{borucki2013} found that the super-Earth is within the HZ of the host star. \citet{kane2014b} arrived at the same conclusion and found (by analysing the HZ boundaries based on stellar parameter uncertainties) that planet "f" is 99.4\% likely to be in the HZ. \citet{kaltenegger2013} assumed, based on the packed system of Kepler-62 with solid planets, that Kepler-62f was formed outside the ice line, indicating water or ice covered surface of the planet depending on the atmospheric pressure of CO$_2$. Based on the assumption that Kepler-62f is indeed a water planet, and using the observed radius, \citet{kaltenegger2013} found that the planet's mass would be $1.1-2.6$ M$_\oplus$. \citet{bolmont2015} used their new \textit{Mercury-T} code to study the evolution of the Kepler-62 system. They found that Kepler-62f potentially have a high obliquity and a fast rotation period, which would result in seasonal effects and both latitudinal and longitudinal winds on the planet. The possible seasonal and latitudinal changes may result in a diverse weather system on the planet, therefore, Kepler-62f may host a quite variable lightning activity.

\item \textbf{Rocky planets with no liquid surface}: These planets supposedly do not have permanent liquid oceans on their surface. However, they still may host a chemically active atmosphere that forms clouds and produces lightning. Lightning production on these planets may also be caused by volcanic activity or electrostatic discharges caused by dust collision (e.g. in dust devils). \citet{schaefer2009} and \citet{miguel2011} modelled different types of potential atmospheres, created by the outgassing of the lava-oceans on the surface of the planet, of hot, volatile-free, rocky super-Earths, and found them to be composed mostly of Na, O, O$_2$, SiO \citep{schaefer2009} and at temperatures $\le 2000$K Fe and Mg \citep{miguel2011}. \citet{ito2015} considered these "mineral atmospheres", evaluated their temperature profiles and investigated their observability via occultation spectroscopy. They considered four rocky planets, CoRoT-7b, Kepler-10b, Kepler-78b, and 55 Cnc e and showed that IR absorption features of K, Na and SiO could be detected in case of Kepler-10b and 55 Cnc e with future missions like the \textit{James Webb Space Telescope}. Such atmospheres would be close to the composition of volcano plumes on Earth and may host lightning activity. We use volcanic lightning flash densities evaluated in Sect. \ref{sec:volc}. The various values in Table \ref{table:vol} (last column) represent various activity stages of eruptions. For example, if we assume that the surface of these planets is covered by almost constantly erupting volcanoes, the flash densities could be very high, like during the phase one of the Mt Redoubt eruption. However, the surface is still covered by volcanoes, but they do not erupt as frequently, or the frequency of explosive eruptions is less, then a smaller flash density can be used, like during the eruptions of Eyjafjallaj\"okull. We also used continental flash density from Earth, though, we note that this value likely underestimates the actual electric activity compared to pure visual inspection of lightning in volcanoes \citep[e.g. Eyjafjallaj\"okull, Sakurojima, Puyehue; see also][]{mcnutt2000}.

\begin{itemize}
\item[$\ast$] Example: \textit{Kepler-10b} \citep[][and number 1 on Fig. \ref{fig:mr}]{batalha2011}.
\end{itemize}

Stellar chromospheric activity measurements (using the Ca II H\&K index) conducted by \citet{dumusque2014} indicate that Kepler-10 is less active than the Sun, which is in accordance with the star's old age (10.6 Gyr). According to \citet{ito2015}, Kepler-10b, a hot, tidally locked rocky super-Earth \citep{dumusque2014}, may host an atmosphere mostly composed of Na, O, O$_2$, SiO and K outgassed from the lava-surface of the planet. The bulk density (Table \ref{table:planet}) of the planet indicates a composition similar to Earth \citep{dumusque2014}.

\begin{itemize}
\item[$\ast$] Example: \textit{55 Cancri e} \citep[][and number 3 on Fig \ref{fig:mr}]{mcarthur2004, vonbraun2011}.
\end{itemize}

Our second candidate for a rocky planet is 55 Cancri e (55 Cnc e), which recently has been reported to be a planet with possible high volcanic activity \citep{demory2015}. The super-Earth orbits the K-type star 55 Cnc on a very close orbit, resulting in a high equilibrium temperature (Table \ref{table:planet}), which may result in the loss of volatiles of the planet. Multiple scenarios have been proposed for its composition including a silicate-rich interior with a water envelope and a carbon-rich interior with no envelope \citep[see][and ref. therein]{demory2015}. A recent study suggests that 55 Cnc e is rather a volcanically very active planet \citep{demory2015}. A large number of volcanic eruptions, especially explosive eruptions, may result in increased lightning activity on the planet due to the large number of volcano plumes. This would allow the production of lightning discharges without the necessity of cloud condensation. \citet{kaltenegger2010} studied the observability of such volcanic activity on Earth-sized and super-Earth-sized exoplanets. They found that large explosive eruptions may produce observable sulphur dioxide in the spectrum of the planet. Similarly to Kepler-10b, 55 Cnc e may host an atmosphere composed of minerals, as a result of the outgassing of the lava on its surface.
 
Combining the findings of studies such as \citet{kaltenegger2010} and observational signatures of lightning, one may confirm a high volcanic activity on terrestrial, close-in exoplanets like Kepler-10b and 55 Cnc e, making these planets interesting candidates for future lightning observations.

\item \textbf{Venus-like planets}: Venus and Earth, though they are similar in size and mass, are very different from each other. Due to Venus' thick atmosphere, the runaway greenhouse effect increases the surface temperature of the planet to uninhabitable ranges. Such exoplanets, Earth- or Super-Earth-size rocky planets with very thick atmospheres, may be quite common \citep{kane2014}. For these planets we can use flash density based on radio observations from Venus. 

\begin{itemize}
\item[$\ast$] Example: \textit{Kepler-69c} \citep[][and number 6 on Fig \ref{fig:mr}]{barclay2013}.
\end{itemize} 

Kepler-69c is a super-Earth sized planet orbiting close to the HZ of a star very similar to our Sun \citep{barclay2013}. \citet{barclay2013} analysed the planet's place in the system and the stellar irradiation and found that Kepler-69c is very close to the HZ of the star or, depending on model parameters, it may lie inside the HZ. They investigated the equilibrium temperature boundaries that Kepler-69c may have, using different albedo assumptions. They found that the temperature of the planet may be low enough to host liquid water on the surface, if not considering an atmosphere. However, a thick atmosphere may increase the temperature high enough (with a low albedo) to prevent water to stay in liquid form \citep{barclay2013}. \citet{kane2013} estimated that most probably Kepler-69c does not lie in the conservative HZ, but rather at a distance equivalent to Venus's distance from the Sun. Also taking into account the stellar flux the planet receives (which is very similar to the incident flux Venus receives ($\sim2600$ W m$^{-2}$)), they defined Kepler-69c as a "super-Venus" rather than a super-Earth \citep{kane2013, kane2014}. The low bulk density calculated by \citet{kane2013} may suggest a silicate and carbonate dominated composition of the planet. In case the planet acquired water during or after its formation, and the evolution of the planet's atmosphere was similar to Venus', then the planet may host a thick CO$_2$ atmosphere \citep{kane2013}. On a Venus-like planet, such as Kepler-69c, lightning activity may be the result of ongoing volcanic activity, or, in the presence of strong atmospheric winds, the electrostatic activity of dust-dust collision.

\item \textbf{Giant gas planets}: In this category we consider planets with sizes (mass and/or radius) in the range of Saturn's to several Jupiter-sizes. Large variety of exoplanets have been discovered, which fall into this category, from close-in hot Jupiters mostly detected by the transit or the radial velocity technique, to young, cool planets hundreds of AU far from their stars detected by direct imaging. We calculate flash densities for the candidate planets based on Saturnian and Jovian flash densities.
	\begin{itemize}
	\item[$\bullet$] \underline{Transiting planets}: Most of the gas giant planets discovered by the transit technique lie within $\sim1.6$ AU from the host star\footnote{Based on data from exoplanet.eu, 29/Jul/2015}. A large number of these planets are found within 0.5 AU, creating a new (not-known from the Solar System) type of exoplanets called "warm-" or "hot-Jupiters", latter ones lying within 0.1 AU \citep{raymond2005}. 

\begin{itemize}
\item[$\ast$] Example: \textit{HD 189733b} \citep[][and number 2 on Fig \ref{fig:mr}]{bouchy2005}.
\end{itemize}
	\end{itemize}
HD 189733 is a K-type star with a hot-Jupiter ("b") in its planetary system. \citet{wright2004} measured Ca H\&K line strength and found the star to be relatively active. Stellar activity due to star-planet interaction has been observed in X-ray \citep[e.g.][]{pillitteri2014} and FUV \citep{pillitteri2015} spectra at certain times of the planetary transit. Namely, after the secondary eclipse, X-ray flares appeared in \textit{XMM-Newton} \citep{pillitteri2014} and \textit{Swift} \citep{lecavelier2012} data, while a brightening in the FUV spectrum was also seen \citep{pillitteri2015}. \citet{pillitteri2015} explained the FUV features by material accreting onto the stellar surface from the planet. \citet{see2015} investigated exoplanetary radio emission variability due to changes in the local stellar magnetic field. They found potential variations up to 3 mJy. The frequency of magnetospheric radio emission \citep[$<40$ MHz,][]{zarka2007} coincides with the radio emission range that lightning may produce \citep[$<\sim100$ MHz,][]{desch2002}. The magnetic radio emission may potentially be a background radio source in lightning radio observations. A slope in the IR transmission spectrum of HD 189733b has been measured by several groups \citep[e.g.][]{pont2008, sing2011}, which was interpreted as a feature caused by cloud-induced Rayleigh scattering in the atmosphere. \citet{mccullough2014} found prominent water features in the NIR transmission spectrum of HD 189733b and simultaneously reinterpreted the slope in the spectrum. They suggest that the slope can be produced by a clear planetary atmosphere and unocculted star spots. \citet{lee2015}, however, support the finding that HD 189733b is covered by a thick layer of clouds. The atmosphere of HD 189733b may host lightning activity due to cloud convection and charge separation due to gravitational settling. This well-studied (see references above) exoplanet is a good candidate for lightning observations, because other effects, like stellar activity, can be modelled easier than for less known systems.

	\begin{itemize}
	\item[$\bullet$] \underline{Directly imaged planets}: Planetary objects detected by direct imaging are way less in number than e.g. transiting exoplanets. These objects, due to the selection effect of the technique, lie far from the host star, from $\sim10$ to thousands of AU. Though we list these objects under the category of gas giant planets we note the ambiguity in the classification due to the uncertainty in the definition of the mass limit between brown dwarfs and planets \citep{perryman2011}. This category may include brown dwarfs, planets, and objects with masses on the borderline \citep[][table 7.6]{perryman2011}. 

\begin{itemize}
\item[$\ast$] Example: \textit{GJ 504b} \citep{kuzuhara2013}.
\end{itemize}
	\end{itemize}

GJ 504, a young, $160^{+350}_{-60}$ Myr old, solar-type star shows X-ray activity typical property of such stars \citep{kuzuhara2013}. \citet{kuzuhara2013} investigated the colour of GJ 504b and found it to be colder (Table \ref{table:planet}) than previously imaged planets. The place of the object on the colour-magnitude diagram suggests that it is rather a late T-type dwarf with a mostly clear atmosphere \citep{kuzuhara2013}. \citet{janson2013} detected strong methane absorption in the atmosphere of GJ 504b, which also indicates that the object is a T-type brown dwarf. Though the atmospheres of these objects are considered to be clear, studies showed that it is possible to form clouds, e.g. made of sulphides \citep{morley2012} in T-type dwarf atmospheres. The potential sulphide clouds make GJ 504b a good candidate of hosting electric discharges in its atmosphere. The object is far enough from the host star so that its internal heating suppresses the external one, which could result in extensive convective patterns, just as lightning hosting clouds may form on Jupiter \citep{baines2007}. Convection and gravitational settling can be viewed as preconditions for lightning to occur.

\item \textbf{Brown dwarfs}: Brown dwarfs have masses from several M$_{\rm Jup}$ (Jupiter mass) to several tens of M$_{\rm Jup}$ and temperatures low enough for cloud formation \citep[e.g.][]{helling2014}. L type brown dwarfs are fully covered by clouds. The variability of L/T transition and most probably T type brown dwarfs is explained by patchy cloud coverage \citep{showman2013, helling2014}. \citet{helling2008} modelled cloud formation on brown dwarfs and derived grain size distributions and chemical composition through the entire atmosphere. They found that the particle size in these atmospheres is of the order of 0.01 $\mu$m (upper layers) to 1000 $\mu$m (deep layers). For a brown dwarf with log $g = 3$ or $5$ and $T_{\rm{eff}} = 1800$ or $1300$ K, the $5 \mu$m particle size range, the assumed thundercloud particle size for Jupiter (see Sect. \ref{subs:jup}), is in the atmospheric layers with $\sim 1300$ K local temperatures, which correspond to the mid layers of the atmosphere \citep[][fig. 4, fourth panel]{helling2008}. They also found the clouds to be made of mixed mineral cloud particles that change their size according to atmospheric height. Volcano plumes, producing lightning flashes, are mostly made of dust. These plumes may resemble brown dwarf clouds. We use flash densities obtained in Sect. \ref{sec:volc}, to estimate lightning occurrence on brown dwarfs. However, these statistics are based on a few eruptions, which does not provide a general idea about volcanic lightning densities, but provide several scenarios with more electrically active and less electrically active dust clouds. We also use flash densities from Jovian and Saturnian thunderclouds, because the particle sizes may resemble brown dwarf dust particles, and basic physical processes of cloud formation are fundamentally same in these environments, though their efficiency may vary, as we discuss it in the following sub-section.

\begin{itemize}
\item[$\ast$] Example: \textit{Luhman 16B} \citep{luhman2013}.
\end{itemize}

Luhman 16B (or WISE J104915.57-531906.1B) is the secondary component of the closest brown dwarf binary system discovered so far, with a distance of 2 pc from the Sun. It is a late L, early T type object representing the L/T transition part of the brown dwarf family \citep{luhman2013}. \citet{crossfield2014} monitored the brown dwarf during its one rotational period \citep[4.9-hour,][]{gillon2013} and mapped its surface using Doppler imaging. They interpreted the revealed features as dust clouds in the atmosphere of the object. \citet{buenzli2015} found the variability of Luhman 16B to be relatively high, up to more than 10\%. Their cloud structure model showed that the variability is caused by varying cloud layers with different thickness, rather than varying cloudy and clear parts of the atmosphere. This means, we see into various levels of cloud regions, making the possibility of detecting lightning inside the atmosphere higher. Similarly to GJ 504b, Luhman 16B may host intensive lightning activity, because cloud formation, convection and gravitational settling determine its atmosphere. Different cloud layers have been detected on Luhman 16b, which may cause similar dynamic structures to occur like on Jupiter and Saturn and may allow the observer to detect lightning inside the atmosphere of the brown dwarf. 

\end{itemize}

\subsection{Flash densities for extrasolar objects} \label{sec:flashdens}

\begin{table}
 \small 
 \caption{Estimated total flash/SED numbers during a transit over the disc of the planet calculated from flash densities in Table \ref{table:plan}. As the values in Table \ref{table:plan} are lower limits, the flash numbers given here represent lower limits too. The bottom four lines of the table present Venus, Earth, Jupiter and Saturn as transiting planets (with inclinations of $90$\textdegree). Transit duration was calculated based on \citet[][equations 6.2 and 6.3]{perryman2011}. Since the determined eccentricity ($e$) for these planetary orbits is low (largest is $\sim 0.1$), for the calculation of the transit time we assumed $e$ to be 0 for all objects. Here we use the full length of the transit (from 1$^{\rm{st}}$ contact to 4$^{\rm{th}}$ contact).}
  \begin{tabular}{@{}lcc@{}}	
	\hline 
	Planet & Transit duration [h] & \vtop{\hbox{\strut Total number of flashes}\hbox{\strut during transit}} \\
	\hline \hline
	Kepler-186f & 6.25 & $4.51 \times 10^5$ \\
	Kepler-62f & 7.72 & $1.34 \times 10^5$ \\
	Kepler-10b & 1.85 & $2.67 \times 10^6$ \\
	55 Cancri e & 1.57 & $4.16 \times 10^6$\\
	Kepler-69c & 11.78 & $3.2 \times 10^{-1}$ \\
	HD 189733b & 1.89 & \vtop{\hbox{\strut $6.57 \times 10^4$}\hbox{\strut (Jupiter)}} \\
	HD 189733b & 1.89 & \vtop{\hbox{\strut $2.04 \times 10^5$}\hbox{\strut (Saturn)}} \\
	\hline
	Venus & 11.15 & $9.0 \times 10^{-2}$ \\
	Earth & 13.11 & $7.67 \times 10^5$\\
	Jupiter & 32.59 & $8.34 \times 10^5$ \\
	Saturn & 43.46 & $2.39 \times 10^6$ \\
	\hline
  \label{table:exo}
  \end{tabular}
\end{table}

\begin{table}
 \small 
 \caption{Estimated total flash numbers during a transit over the disc of Kepler-10b and 55 Cancri e calculated from volcanic flash densities in Table \ref{table:vol} (used flash densities are marked with the row number). See also the caption of Table \ref{table:exo}.}
  \begin{tabular}{@{}lccc@{}}	
	\hline 
	Planet & \vtop{\hbox{\strut Transit duration}\hbox{\strut [h]}} & \vtop{\hbox{\strut Total number of}\hbox{\strut flashes during transit}} & \vtop{\hbox{\strut Row,}\hbox{\strut Table \ref{table:vol}}} \\
	\hline \hline
	\multirow{5}{*}{Kepler-10b} & \multirow{5}{*}{1.85} & $1.02 \times 10^8$ & $[1]$ \\
	 & & $3.26 \times 10^8$ & $[2]$ \\
	 & & $1.23 \times 10^{10}$ & $[3]$ \\
	 & & $2.04 \times 10^{12}$ & $[4]$ \\
	 & & $1.11 \times 10^{11}$ & $[5]$ \\
	\hline
	\multirow{5}{*}{55 Cancri e} & \multirow{5}{*}{1.57} & $1.59 \times 10^8$ & $[1]$  \\
	 & & $5.07 \times 10^8$ & $[2]$ \\
	 & & $1.91 \times 10^{10}$ & $[3]$ \\
	 & & $3.17 \times 10^{12}$ & $[4]$ \\
	 & & $1.73 \times 10^{11}$ & $[5]$ \\
	\hline
  \label{table:exovol}
  \end{tabular}
\end{table}

Table \ref{table:plan} lists extrasolar objects with their Solar System counterparts. Based on the data available, we arrange these objects into six groups (Sect. \ref{sec:exopl}). From Earth we obtained three flash densities using LIS/OTD flash observations and WWLLN and STARNET sferics detections. Strokes detected by WWLLN and STARNET were converted to flashes \citep[assuming 1.5 sferics/flash;][]{rudlosky2013}. We assumed that a planet with a similar surface to Earth's, in the HZ of the star has the same flash density, as the global value on Earth. Kepler-186f is our candidate for these conditions. However, depending on the continent-ocean fraction and the amount of insolation of the planet, the flash density may vary. We consider Kepler-10b and 55 Cnc e to be rocky planets with no liquid surface. Although it is arguable whether these planets host an atmosphere, in case they do, lightning activity may be similar to the activity over Earth-continents. Both planets may also be good candidates for volcanically active planets, resulting in lightning discharges in volcano plumes. Similarly, we used flash densities from oceanic regions in order to simulate lightning statistics on Kepler-62f, a presumed ocean planet. Earth is the most well studied planet, resulting in the most accurate flash densities obtained. Uncertainties raise, however, from the accuracy with which we can determine the similarities between the exoplanet and Earth or a Terran environment. The arguments for our approach, such as similarities between Earth and the exoplanets in size, composition or cloud occurrence, are discussed in Sections \ref{sec:exopl} and \ref{sec:casest}.

From an astrophysical perspective, Jupiter and Saturn have the same flash densities within an order of magnitude (Table \ref{table:plan}). Two types of gas giant planets are studied, HD 189733b a hot Jupiter, and GJ 504b a fairly cold giant planet in the outer regions of the stellar system, which has been suggested to be comparable to a brown dwarf of spectral type T. These planets represent the two edges of giant planetary bodies, the former being a highly insolated one, while for the latter internal heating has a higher contribution to global temperatures and cloud formation. For Luhman 16B, representing the L/T transition brown dwarfs with most probably patchy cloud coverage, Jupiter was considered as a good analogue. The flash densities obtained for Jupiter and Saturn carry relative large uncertainties, due to the fact that these planets are much less studied than Earth. The observations have been carried out for less time with less frequency, with less sensitive instruments. However, the flash densities listed in Table \ref{table:plan} serve well as lower statistical limits for these planets and their extrasolar counterparts. We support our approach of using Solar System, in this case Jovian and Saturnian, lightning statistics as guidance for extrasolar studies, with the fact that the basic physical processes of cloud formation are fundamentally the same in every environment, though their efficiency may vary. It also has been shown that the electric field breakdown initializing a lightning discharge does not depend strongly on the chemical composition of the gas \citep{helling2013}. Charging mechanisms necessary for the build up of the electric filed have been investigated and, e.g., \citet{rodriguezbarrera2015} showed that, for an atmosphere to be ionized, just thermal ionization is enough to produce large amount of positive and negative charges. Therefore, we suggest that Solar System lightning flash densities provide good first estimates for extrasolar lightning occurrence.

Table \ref{table:vol} lists flash densities of various eruptions of two volcanoes, Eyjafjallaj\"okull and Mt Redoubt. We suggest that these flash rates may resemble several scenarios on rocky exoplanets with no water surfaces (Kepler-10b, 55 Cnc e) and on brown dwarfs (Luhman 16B). Such scenarios may include a surface fully covered by volcanoes erupting very frequently. In this case flash densities may be as high as it was during the first phase \citep["explosive phase",][]{behnke2013} of the 29 March 2009 eruption of Mt Redoubt. Volcanically very active surfaces, but with not that frequent explosive eruptions, may have flash densities of the order of the Eyjafjallaj\"okull values. Dust charging in brown dwarf atmospheres may be similar to charging in volcano plumes, and could produce flash densities similar to Eyjafjallaj\"okull densities and the values of the second phase \citep["plume phase",][]{behnke2013} of the 29 March 2009 Mt Redoubt eruption.
 
The majority of extrasolar planets was discovered by the transit method\footnote{http://exoplanet.eu/ (on 26/01/2016)}. Transit observations and measurements taken during the transit or the occultation of the planet are the most successful techniques in characterizing exoplanets and their atmospheres. Therefore, it is interesting and informative to see how much lightning could be observable during a planet's transit. This information will further allow us to determine observable signatures of lightning coming from these planets. Also, it is a good example to show how the obtained hourly flash densities of this paper can be used for scientific predictions. Table \ref{table:exo} lists the transiting exoplanets introduced in the previous sections and Venus, Earth, Jupiter and Saturn as a transiting planet. Table \ref{table:exovol} lists flash densities calculated for two transiting planets, Kepler-10b and 55 Cnc e, based on volcanic lightning densities listed in Table \ref{table:vol}. These tables summarize how many flashes could be present during a transit on the disc of the planet observed from 1$^{\rm{st}}$ to 4$^{\rm{th}}$ contact \citep{perryman2011}. The projected surface area (disc) of a planet is given by $2r^2\pi$, where $r$ is the mean radius of the planet. The total number of flashes during the transit is calculated from the hourly flash densities (flashes km$^{-2}$ h$^{-1}$) given in Tables \ref{table:plan} and \ref{table:exovol} by multiplying these values by the area (km$^2$) of the planetary disc and the length of the transit (h). The gas giant, HD 189733b is listed twice in Table \ref{table:exo} indicating that flash densities from both Jupiter and Saturn (both averaged from the values in Table \ref{table:plan}) were used to estimate lightning occurrence on this planet. Since the two Solar System planets show similar densities (Table \ref{table:plan}), we obtain similar results for HD 189733b when using the flash densities from Jupiter and Saturn. If HD 189733b should develop a storm feature similar to Saturn's gigantic 2010/11 storm with its extremely high flash density, then  this might also produce potentially observable signatures on HD 189733b even during its short transit time. Comparing the values for Kepler-10b and 55 Cnc e in Tables \ref{table:exo} and \ref{table:exovol}, it is clearly seen that volcano eruptions produce much higher lightning activity, than thunderclouds. However, it is important to note that, while we took average values for continental thundercloud lightning activity, we have extreme values from certain eruptions that might not resemble average flash densities of volcanic plumes. The numbers listed in the third column of Tables \ref{table:exo} and \ref{table:exovol} are guides to a lightning flash count we can expect for the listed transiting planets. The flash densities shown in Tables \ref{table:exo} and \ref{table:exovol} suggest a relatively high lightning activity on these planets when they are observed during a full transit. This increases the probability of measuring signals resulting from lightning discharges. However, further study is necessary to estimate the energy content of the discharges on each planet, and so observability of their signatures. 

For example, let us consider HD 189733b. During its 2-hour long transit, about $10^{5}$ lightning flashes occur on the projected surface according to our calculations. If we assume that the average total energy content of these flashes is $\sim 10^{12}$ J \citep[p. 43]{leblanc2008}, based on Jovian lightning optical efficiency calculations \citep{borucki1987}, then the total energy dissipated from lightning discharges during the transit of HD 189733b is of the order of $10^{17}$ J, or $10^{5}$ TJ. For comparison, on Earth a typical lightning flash releases energy of the order of $10^{9}$ J \citep{maggio2009}. Energy measurements of Earth lightning suggest that about $1-10$\% of the total energy is released in optical \citep[][p. 334]{borucki1987, hill1979, lewis1984} and $\sim 1$\% in the radio \citep{volland1984, farrell2007}. This leaves us with a bit less than 90\% of energy going into mechanical and thermal release, affecting the local chemistry of the atmosphere, which will produce yet unexplored observable spectral signatures. Going back to our example, during the transit of HD 189733b, $9 \times 10^{4}$ TJ energy would affect the atmosphere of the planet. This example benefits from previous lightning energy estimates; however, these estimates are based on Earth lightning properties. Once the energy release from lightning in various extrasolar planetary atmospheres is studied, one can estimate, based on our lightning climatology statistics, how much energy is released not just into observables (optical and radio emission) but to energy affecting the local chemistry. This energy and the caused chemical changes can be further explored and determined whether it is enough to produce observable emission lines in the spectrum of the planet, or the stellar light and planetary thermal emission would suppress these transient signatures.

\subsection{Observational challenges: Effects of stellar activity} \label{sec:stelact}

Apart from technical issues (such as instrumental limits, detection thresholds, etc.), there are natural effects limiting observations, mostly coming from stellar activity. Cool dwarf stars, G K and M spectral types, are in more favour of exoplanet surveys, than hotter ones. G and K stars are the targets of scientists looking for an Earth twin orbiting a Sun-like star. M dwarfs, apart from being the most widespread stars in the Galaxy, are small stars making it easier to detect variation caused by planets in their light curves (the planet-star size ratio can be large enough to detect small planets around the M dwarf) or spectra (variations caused by smaller planets around a small star can be detected easier). However, \citet{vidotto2013} and \citet{see2014} have shown that M dwarfs might not be as good candidates for the search of Earth-like habitable planets because of their high stellar activity. This activity significantly reduces the size of a planetary magnetosphere exposing the planetary atmosphere to erosive effects of the stellar wind. G and K stars have similar activity cycles to the Sun's (11-year cycle), younger stars being rapidly rotating and more active, than older ones \citep{baliunas1995}. Early M dwarfs (M3 and earlier) have radiative cores and outer convective zones indicating similar dynamo processes to the Sun's \citep{west2008}. Later type M dwarfs are fully convective, therefore no solar-like dynamo can operate in them, which result in the change of magnetic field structure \citep{donati2009}. Later M dwarfs in general are more active than earlier type ones, keeping their activity for longer, probably due to this change in magnetic field structure and rapid rotation. As stars age, their rotation slows down and they become close to inactive \citep{west2008}.

However, the activity of the star may support lightning activity in close-in planets. Studies suggest a correlation between solar activity and the number of days with thunderclouds. \citet{pintoneto2013} analysed data of a $\sim 60$-year period in Brazil looking for 11-year cycle variations in thunderstorm activity correlated to solar activity. They suggested that the anti-correlation they found is the result of solar magnetic shielding of galactic cosmic rays, which are thought to have a large effect on lightning production. \citet{scott2014} found a correlation between the arrival of high-speed solar wind streams at Earth, following an increase in sunspot number and decrease in solar irradiance, and lightning activity. They measured the correlation based on lightning occurrence over the United Kingdom using UK Met Office radio observations. This correlation may be the result of increasing number of solar energetic particles reaching the upper atmosphere (coming from the solar wind), which triggers discharges and may increase the number of lightning events. \citet{siingh2011} compared different studies (Brazil, USA, India) and concluded that the relation between lightning activity and sunspot numbers is complex, since data showed correlation in the USA and Brazil and anti-correlation in the Indian Peninsular \citep[see][fig. 6]{siingh2011}. The STARNET and WWLLN data analysed in our study also show more lightning from 2013, close to solar maximum, than from 2009, when the Sun was at its minimum of activity (Fig. \ref{fig:4}). However, \citet{rudlosky2013} showed an improvement of 10\% of the WWLLN DE between 2009 and 2013, which may also be the cause of more stroke detections in 2013.

\section{Summary} \label{sec:con}

The presently known ensemble of exoplanets is extremely diverse, including Earth-like planets and giant gas planets some of which resemble brown dwarfs. A large number of these objects have atmospheres where clouds form. Clouds are known to discharge their electrostatic energy in the form of lightning.
Discharge processes have an effect on the local environment, creating non-equilibrium species.

This paper uses Solar System lightning statistics for a first exploratory study of potential lightning activity on exoplanets and brown dwarfs. We present lightning flash densities for Venus, Earth, Jupiter and Saturn, based on optical and/or radio measurements. We also include lightning- and lightning energy- distribution maps for the gas giant planets and Earth. The obtained information in Sections \ref{sec:earth} and \ref{sec:solsys} was used to estimate lightning occurrence on extrasolar planetary objects. Our sample of extrasolar objects contains transiting planets (Kepler-186f, Kepler-62f, Kepler-10b, 55 Cancri e, Kepler-69c and HD 189733b), directly imaged planets (GJ 504b) and brown dwarfs (Luhman 16B). Transmission spectra are relatively easy to be taken and may contain signatures of lightning activity. Directly imaged planets are far enough from their parent star to be observed directly, and the effects of stellar activity are less prominent, such as in the case of non-irradiated brown dwarfs. Brown dwarfs, because they are close to us, are one of the most promising candidates for lightning-hunting. We defined six categories of extrasolar bodies, with one or two examples, in analogy to Solar System planets. All of these candidates potentially host an atmosphere with clouds, based on either observations or atmospheric models (Sect. \ref{sec:casest}). These examples were chosen because they have common features with Solar System planets or lightning hosting environments (e.g. Kepler-62f being an ocean planet, 55 Cnc e hosting extreme volcanic activity, etc.), or because they represent a specific object type, such as hot Jupiters (HD189733b), Jupiter-sized planets at large distances from the star (GJ 504b), or brown dwarfs (Luhman 16B), which also have a great potential for lightning activity \citep{helling2013, bailey2014}. We suggest that these exoplanets could be potential candidates for lightning activity in their atmospheres based on their characteristics and on our knowledge on lightning forming environments. However, we note that the flash densities estimated in this study are affected by several uncertainties, mostly due to instrumental limits and, in case of Jupiter and Saturn, the lack of temporally and spatially extensive data sets. Regardless, the obtained flash densities give a first guidance for the study of extrasolar lightning.

We had the best data coverage from Earth (using data from the LIS/OTD optical satellites, and the STARNET and WWLLN radio networks), which resulted in more accurate flash densities than from the other planets. Earth provided us with three different options: for ocean planets flash densities from over the Pacific ocean were used, for rocky planets with no water surface, where mineral clouds may form as was shown both by models \citep[e.g.][]{helling2008b, helling2008, miguel2011, lee2015} and observations \citep[e.g.][]{kreidberg2014, sing2009, sing2015}, values from over continents were used, while we considered Earth-twins with similar continent/ocean coverage and with a global flash density from Earth. Data for Jupiter and Saturn were taken from published papers, these include \textit{Galileo} \citep{little1999, dyudina2004}, \textit{New Horizons} (Jupiter) \citep{baines2007} and \textit{Cassini} (Saturn) \citep{dyudina2013} observations. The derived flash densities were used to represent giant gas planets and brown dwarfs. The special case of Venus (only whistler observation with no coordinates for flashes) allowed us to estimate flash densities but not to create a lightning climatology map as it was done for the three other Solar System planets (Figures \ref{fig:1}-\ref{fig:4}, \ref{fig:jup}, and \ref{fig:7}). We also considered volcanic lightning flash densities in case of Kepler-10b, 55 Cnc e and Luhman 16B. These densities are guides for special scenarios discussed in the previous sections.

Table \ref{table:plan} summarizes our findings of planetary flash densities, while \ref{table:vol} shows flash densities of example volcano eruptions. All numbers are expected to be higher because the guiding data provide lower limits as only the most powerful events in the optical and radio wavelengths are detected. No other spectral energies were taken into account here. Using these flash densities, we estimated the global or regional distribution of lightning in space and time. Most of the planets listed under the defined categories are transiting objects, with the potential of taking their transmission spectra, hence possibly observing lightning spectral features. In Tables \ref{table:exo} and \ref{table:exovol} we list the total number of flashes that might occur on these planets during their full transit. We find that volcanically very active planets would show the largest lightning flash densities if lightning occurred at the same rate on these planets as it does in volcano plumes on Earth. It is also prominent that the exoplanet HD 189733b would produce high lightning occurrence even during its short transit, if it had a large storm occurring in its atmosphere, like the one on Saturn in 2010/11. If we knew how energetic lightning was on exoplanets, we could estimate the total energy released from lightning flashes inside the atmosphere based on the findings in Tables \ref{table:vol} and \ref{table:plan}, and consider whether signatures of these flashes would be detectable. Depending on the chemical composition of the atmosphere, various emission lines can appear in the spectrum of lightning \citep[e.g.][]{wallace1964, orville1965, weidman1989, borucki1996, krasnopolsky2006, bailey2014, xue2015}. As the planet orbits the star, the emission lines will be Doppler-shifted. Cross-correlating the observed dayside spectra of the planet with known lightning spectra, we can, in principle, use the planet's orbital motion to check whether its spectrum contains lightning spectral features. The cross-correlation technique \citep{snellen2010, brogi2013} has been used to observe the molecular content of planetary atmospheres. \citet{brogi2012} detected CO absorption in the dayside spectrum of the non-transiting planet, $\tau$Bo\"otis b, by tacking high resolution spectra through three days, mapping about $1/4$ of the planet's orbit. In an ongoing project we study lightning properties in extrasolar environments and the observability of signatures produced by such lightning discharges \citep[e.g.][]{hodosan2016}. This topic requires an extended study and is beyond the scope of the current paper.


\section*{Acknowledgements}

We thank Carlos Augusto Morales Rodrigues from STARNET, Robert H. Holzworth from WWLLN and Daniel J. Cecil from LIS/OTD for their kind help in obtaining data from the lightning detecting networks and satellites. We wish to thank the World Wide Lightning Location Network (http://wwlln.net), a collaboration among over 50 universities and institutions, for providing the lightning location data used in this paper. We thank Richard A. Hart for his kind help with the \textit{Venus Express} data. We thank Yoav Yair, Craig R. Stark, Andrew Collier Cameron and Victor See for helpful discussions. We highlight financial support of the European Community under the FP7 by an ERC starting grant number 257431. R.A.T. thanks the Royal Astronomical Society (RAS) and the Physics Trust of the University of St Andrews for supporting his summer placement at the University of St Andrews.

\bibliographystyle{mnras}
\bibliography{bib}

\appendix

\section[]{Summary of surveys}


\subsection{Lightning detection in the optical: Optical Transient Detector (OTD)/Lightning Imaging Sensor (LIS)} \label{subs:e_op}

Both the Optical Transient Detector (OTD) and the Lightning Imaging Sensor (LIS) are space-born instruments dedicated to scan the atmosphere of the Earth for quickly-varying phenomena such as lightning. OTD was in operation between 1995 and 2000 on board of the Microlab-1 (OV-1) satellite orbiting 735 km above the terrestrial surface on an orbit with inclination of $70$\textdegree with respect to the Equator, allowing the monitoring of the whole globe, but excluding the polar regions \citep{boccippio2000}. LIS was in operation between 1997 and 2015 on board of the Tropical Rainfall Measuring Mission (TRMM)\footnote{http://trmm.gsfc.nasa.gov/}. The satellite's orbit was restricted to the tropical region, between $\pm 38$\textdegree in latitude, 350 km above the Earth \citep{beirle2014}. Both OTD and LIS detect lightning flashes by monitoring the 774 nm oxygen line in the lightning spectrum \citep{beirle2014}. The optical observations allow the detection of CG, IC and cloud-to-cloud discharges from space. 

The composite, gridded data set of OTD/LIS gives information on the location and time of occurrence of individual flashes, including the number of events (pixels exceeding the intensity background threshold) and groups (events occurring in adjacent pixels within the same integration time) that the flashes (groups occurring within 330 ms and within 15.5(OTD)/6.5(LIS) km) are composed of \citep{beirle2014}. The OTD/LIS data used here were obtained on 18 July 2014\footnote{http://thunder.nsstc.nasa.gov/data/data\_lis-otd-climatology.html} (Daniel Cecil (private com.)) for the period of $1995-2013$. The downloaded data include different types of flash rates such as mean annual flash rates, annual cycles of flash rates and daily time series of flash rate, raw flash counts and flash counts scaled by detection efficiency.


\subsection{Instruments for detection of lightning in the radio frequency range: STARNET and WWLLN} \label{subs:rad}

The Sferics Timing and Ranging Network (STARNET) is an Earth-based radio network currently composed of 11 antennas operating in the very low frequency (VLF) range ($7 - 15$ kHz). STARNET has been in operation since 2003 in Africa and since 2006 in Brazil and the Caribbean (previously operating as a test network in the United States between 1993-1998). STARNET has integrated the European ZEUS lightning network, which was operated until 2005\footnote{http://www.zeus.iag.usp.br/index.php?lan=en}.

The publicly available STARNET data is composed of monthly and daily processed sferic information including the time of observation (date and time to milliseconds), location of the origin (latitude, longitude), arrival time difference (ATD) error in ms and quality control. The position of the sferics is determined by using the ATD technique that involves the measurement of the time difference between the detection of the individual sferics with different antennas. For this technique to work, at least four antennas have to observe the radio signal \citep{morales2014}. We obtained data on 17 July 2014\footnote{http:/www.zeus.iag.usp.br/index.php? lan=en} (Carlos Augusto Morales Rodrigues (private com.)) for years 2009 and 2013. STARNET detects individual strokes in the flashes (flashes are composed of multiple strokes as shown on \citet[Fig 4.1]{rakov2003}).

The World Wide Location Network (WWLLN) is a developing lightning location network that observes VLF ($3 - 30$ kHz) sferics. WWLLN currently includes $\sim 70$ stations all around the world \citep{hutchins2013}. It detects both IC and CG discharges (individual strokes in flashes) but is more sensitive to the CG flashes since they are stronger than the IC ones \citep{rudlosky2013}.  

WWLLN data were obtained\footnote{http://www.wwlln.com/} (Robert H. Holzworth (private com.)) in Aug-Sep 2014 for the years 2009 and  2013, however, there are 15 days missing from the 2009 series (first part of April). The data files include, amongst others, locations (latitude, longitude) of strokes, times of observations and energy estimates. A separate file contains the relative detection efficiencies in the form of maps for each hour of each day.

\section[]{Lightning flash location errors in the Galileo, New Horizons and Cassini data} \label{app:b}

In this section we explain in detail how the corrections for lightning flash locations, introduced as "error bars" or "errors", shown in Fig. \ref{fig:grid} and mentioned in Sect. \ref{subs:climatjs} were calculated. These corrections are applied when plotting lightning flash appearances in Figures \ref{fig:jup} and \ref{fig:7}. The reason for this application is that the observing instruments have pointing errors, which results in an uncertainty of the location measurement for lightning flashes.

Galileo latitude correction from \citet{little1999}:

\begin{equation} \label{aeq:1}
{\rm error} = \frac{{\rm point}_{\rm err}}{\cos({\rm lat})} \times \frac{360}{462000},
\end{equation}

\noindent where point$_{\rm err} = {\rm m} \times {\rm res}$ is the pointing error in km, m is 20 pixels or 40 pixels depending on the observing mode \citep{little1999}, res is the image resolution in km, lat is the latitude at which the lightning flash was observed, and $360/462000$ converts km to degrees. The first term of Eq. \ref{aeq:1} (${\rm point}_{\rm err} / \cos({\rm lat})$) is the spatial resolution of the instrument.

Similarly, for the New Horizons and the Cassini data, spatial resolution given in km is taken from \citet{baines2007} and \citet[][Supplement]{dyudina2013} and it is converted into degrees to get the latitude correction. In case of Saturn and the Cassini data, the conversion factor is $360/378680$.

\section[]{Tables}

\begin{table*} 
\begin{threeparttable}
 \small 
 \centering
 \caption{Stellar properties of the stars hosting the example exoplanets listed in this paper.}
  \begin{tabular}{@{}llllll@{}}	
	\hline 
	Star & Spectral type & \vtop{\hbox{\strut Effective Temperature}\hbox{\strut estimate (T$_{\rm eff}$/K)}} & \vtop{\hbox{\strut Mass}\hbox{\strut (M$_*$/M$_\odot$)}} & \vtop{\hbox{\strut Radius}\hbox{\strut (R$_*$/R$_\odot$)}} & Reference \\
	\hline	
	Kepler-186 & M1V\tnote{(1)} & $3790$ & $0.478 \pm 0.055$ & $0.472 \pm 0.052$ & \citet[][Supplement]{quintana2014} \\
	Kepler-62 & K2V & $4930$ & $0.69 \pm 0.02$ & $0.64 \pm 0.02$ & \citet[][Supplement]{borucki2013} \\
	Kepler-10 & G & $5710$ & $0.910 \pm 0.021$ & $1.065 \pm 0.009$ & \citet{dumusque2014} \\
	55 Cnc & K0IV-V & $5200$ & $0.905 \pm 0.015$ & $0.943 \pm 0.01$ & \citet{vonbraun2011} \\
	Kepler-69 & G4V & $5640$ & $0.81^{+0.09}_{-0.081}$ & $0.93^{+0.18}_ {-0.12}$ & \citet{barclay2013} \\
	HD 189733 & K2V\tnote{(2)} & $5050$ & $0.82 \pm 0.03$ & $0.76 \pm 0.01$ & \citet{bouchy2005} \\
	GJ 504 & G0V & $6230$ & $1.22 \pm 0.08$ & & \citet{kuzuhara2013} \\
	\hline
  \label{table:star}
  \end{tabular}
  \begin{tablenotes}
	\item[1] http://simbad.u-strasbg.fr/simbad/sim-basic?Ident=kepler-186
	\item[2] \citet{fares2013}        
  \end{tablenotes}
\end{threeparttable}
\end{table*}

\bsp	
\label{lastpage}
\end{document}